\documentclass[twocolumn,numberedappendix,iop]{emulateapj}

\usepackage{amssymb}
\usepackage{amsmath}
\usepackage{multirow}
\usepackage{wasysym}

\shorttitle{Magnetic Grain Trapping}
\shortauthors{G.\ H.\ Rieke et al.}

\begin{document}

\title{Magnetic grain trapping and the hot excesses around early-type stars}

\author{G.\ H.\ Rieke}
\author{Andr\'as G\'asp\'ar} 
\author{N.\ P.\ Ballering}
\affil{Steward Observatory, University of Arizona, Tucson, AZ, 85721\\
grieke@as.arizona.edu, agaspar@as.arizona.edu, ballerin@email.arizona.edu}

\begin{abstract}

A significant fraction of main sequence stars observed interferometrically in the near infrared have slightly extended components 
that have been attributed to very hot dust. To match the spectrum appears to require the presence of large numbers 
of very small ($<$ 200 nm in radius) dust grains. However, particularly for the hotter stars, it has been unclear how 
such grains can be retained close to the star against radiation pressure force. We find that the expected 
weak stellar magnetic fields are sufficient to trap nm-sized dust grains in epicyclic orbits for a few weeks or longer, sufficient 
to account for the hot excess emission. Our models provide a natural explanation for the requirement that the hot excess dust grains be smaller than 200 nm. They also suggest that magnetic trapping is more effective for 
rapidly rotating stars, consistent with the average vsini measurements of stars with hot excesses being larger 
(at $\sim 2 \sigma$) than those for stars without such excesses. 

\end{abstract}

\section{Introduction}

Planetary debris disks consist of dust and larger grains produced by planetesimal collisions, events that are driven 
through gravitational shepherding and stirring by planets \citep[e.g.,][]{liou1999,wyatt2008}. One might therefore 
expect the structure of debris disks to be so strongly affected by the placement of the planets and planetesimal belts 
that each system would be completely unique. However, it appears that there is a general regularity to debris disk 
structure \citep{su2014}, with a very dilute terrestrial planet zone often marked by silicate emission, a denser zone 
roughly corresponding to our asteroid belt and lying near the snow line, an outer zone analogous to the Kuiper Belt, 
and a dilute halo consisting of small grains on highly elliptical orbits analogous to $\beta$ meteoroids and in many 
cases also grains being ejected from the system by radiation pressure force. In addition, a fifth zone of hot emission 
predominantly around A-type stars (although found around some stars of later type) has been 
revealed by near-infrared interferometry near 2 $\mu$m \citep{absil2013, ertel2014}. 

The hot component was an unexpected discovery, and a number of possible explanations were explored to explain it 
\citep[e.g.,][]{absil2008, akeson2009}, including (1) close companions; (2) stellar oblateness; (3) scattering of 
stellar emission by dust; (4) emission by optically thin gas; and (5) thermal emission by very hot dust. Although the 
first four may apply in a minority of cases, only the fifth appears capable of accounting for the hot emission in 
the majority of the stars where it is detected.  \citet{vanlieshout2014} analyzed the ability 
of dust produced in conventional debris disk models to 
explain the hot emission. In these models, larger bodies are continuously ground down into smaller ones and dust by a 
cascade of collisions. The orbits of the small particles experience a number of forces. Stellar radiation provides a 
radial outward force and also an orbital-velocity-vector-dependent force, also known as Poynting-Robertson drag (PRD), 
which results in a slow inward circularized spiraling of a particle\footnote{Stellar winds can produce forces analogous 
to those from stellar radiation, imposed by protons originating from the central star.}. The inward spiral 
gradually brings dust close to the star where it can become very hot; however, once the products of the collisions become 
small enough (sizes of order 3 - 7 $\mu$m around A-stars), they are ejected by radiation pressure force. Based on extensive 
analytic and numerical modeling of a specific case in which PRD brought grains inward from an outer ring, 
\citet{vanlieshout2014} showed that this conventional approach is unable to produce the spectral energy distributions 
(SEDs) and brightnesses at 2 $\mu$m characteristic of the hot excesses.   

Generically, given the detections and detection limits at 2 and 10 $\mu$m \citep{absil2013,mennesson2014, ertel2014}, 
producing the hot excesses requires a large population of sub-micron-sized dust particles, which under conventional debris 
disk models are quickly expelled from the vicinity of A-stars by radiation pressure force 
\citep[e.g.,][]{defrere2011, lebreton2013}. To date, the problem of retaining these small grains has not been solved 
\citep[e.g.,][]{bonsor2013}, leading to models for enhanced inward planetesimal transport that are both very specific 
and quite complex \citep{raymond2014,bonsor2014} and thus difficult to picture around {\it all} the stars with hot 
emission.

This paper examines a number of aspects of the hot excesses, which in accordance with previous work we will assume arise 
through thermal radiation by small dust grains. We focus on hot excesses around A-stars, since this situation is most 
challenging to explain (given the strong radiation pressure force on small dust grains). We will address later-type stars 
briefly after the analysis of A-star excesses. In Section 2 we analyze near-infrared photometry to help confirm the reality of 
the hot excess emission independently of interferometry. We then explore the parameters required for hot dust to create this 
emission in Section 3. In Section 4, we show that the dust can be trapped by the stellar magnetic field to 
increase its dwell time near the star and thus to enhance its emission. Section 5 considers the effect of stellar type, magnetic field strength, and 
rotation on this behavior. Section 6 discusses the mass transport and equilibrium density of nanograins to compare 
with the requirements to create the hot emission. Section \ref{sec:comets} describes how infalling comets and asteroids
can enhance the mass transport where that is required to match the intensity of the hot excess. Section \ref{sec:tests} 
tests some predictions of our model, and Section \ref{sec:end} summarises our results. Appendix A provides details of the 
photometry used in Section 2. 

\section{A test for the presence of hot dust debris}
\label{sec:hottest}

Slightly extended emission has been detected around many stars through interferometry at 2 $\mu$m (see summary in 
\citet{absil2013}), and around some at 1.6 $\mu$m \citep{ertel2014} and at 10 $\mu$m \citep{mennesson2014}, through the 
reduced visibility compared with point reference sources. Given that these measurements are nearly all close to the 
limiting capability of the instruments, we report here an effort to confirm this effect using photometry. We will 
focus on the $J - K_S$ colors of stars of spectral types F6V and earlier. This range of spectral type is selected because 
later types have extreme sensitivity in this color to small errors in spectral type and/or stellar temperature, as well 
as to spectral absorption lines. This sensitivity makes the uncertainties in the $J - K_S$ color resulting from errors 
in stellar temperature too large for our purposes.

\begin{figure}
\begin{center}  
\includegraphics[angle=0,scale=0.3]{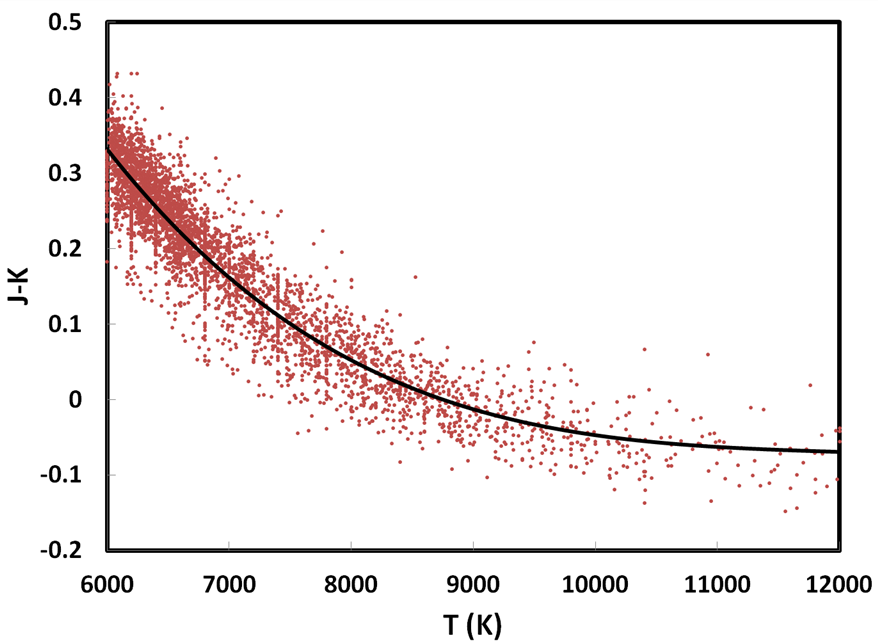}
\caption{J-K color vs. stellar temperature and the fourth-order fit used to represent the trend.}
\label{fig:J-K}
\end{center}
\end{figure}

To begin, we determined the trend of $J - K_S$ color with stellar temperature. This color was selected to minimize 
systematic errors between different photometric systems (JHK usually being measured in a single system) and because the 
sublimation temperature of carbon grains suggests that the contrast of the hot dust emission relative to the star should 
be $\sim$ five times lower at $J$ than at $K_S$. We obtained 2MASS and visible \citep{pickles2010} photometry for the 
Hipparcos stars listed by \citet{mcdonald2012}. To have a uniform set of temperature estimates, we used the photometric 
approach of \citet{casagrande2010}. We utilized the relation for temperature on the basis of $V - K_S$ rather than those 
at shorter wavelengths because the long wavelength baseline and use of infrared bands should make the results less affected 
by interstellar reddening. We then fitted a 4th order polynomial to the trend of $J - K_S$ with temperature, and clipped 
outliers from successive fits down to the level of $\pm$ 0.15 mag. The survivors, 97.4\% of the original sample, provided 
the final fit (Figure \ref{fig:J-K}). 

We next obtained $J$ and $K_S$ photometry for as many stars as possible from the near-infrared interferometric programs 
\citep{absil2013,ertel2014}, requiring spectral types of F6 or earlier and a consensus in the spectral type measurements 
\citep{skiff2014} that the star is not a giant. Many of the stars are too bright for accurate modern photometry 
(e.g., in 2MASS) so we transformed heritage photometry into the 2MASS system as described in Appendix A. We compared the 
observed (and transformed) $J - K_S$ colors of all the stars within our spectral type criterion and with high quality 
photometry with expectations from the fit of the run of this color vs.\ temperature. 

There were 41 stars altogether from \citet{absil2013} and \citet{ertel2014} that met all of our requirements. The rms 
scatter for these stars around the expected $J - K_S$ was 0.017 mag. We found that the stars with identified hot 
excesses\footnote{HIP 27321, 57632, 58803, 60965, 91262, 92043, 93747, 97640, 105199, and 109422. HIP 109857 was 
eliminated because its excess is from a faint companion \citep{mawet2011} and HIP 28103 was not included because it is an X-ray source \citep{voges1999}, 
indicating it also probably has a faint red companion star \citep{derosa2011}.} have larger $J - K_S$ than the rest of the sample\footnote{HIP 2802, 11001, 11783, 16245, 21589, 
22499, 27288, 32362, 41307, 50191, 53910, 54872, 59774, 61622, 67275, 71284, 71908, 78702, 73996, 74824, 76829, 86486, 
87108, 88175, 88771, 94114, 98495,  102333,  102485, 109285, and 111449} by $0.009 \pm 0.006$ mag (error of the mean). 
However, if the small additional red color is real, the temperatures of these stars from $V - K_S$ will be made to appear 
slightly cooler than they should be intrinsically. To estimate this effect, we corrected $V - K_S$ for every star with an 
identified excess downward by 0.01 mag and repeated the comparison, finding a corrected difference in $J - K_S$ of 
$0.011 \pm 0.007$. The redder color relative to the fit of color vs.\ temperature is significant at about 1.5 $\sigma$, 
or equivalently there is only about a 6\% probability of such a color being obtained by chance. Vega was not included in 
our calculations. It is discussed by \citet{rieke2008}, where it is shown that the contribution from the hot excess is 
of significant help in reconciling the observed colors with expectations; the photometry of this star alone supports the 
reality of the hot excess at the $\sim$ 1 $\sigma$ level. 

To check against possible systematic errors due to the temperature estimates for the stars, we repeated this study using 
the temperatures from fits to the stellar SEDs in \citet{mcdonald2012}. The outcome was essentially identical - a 
redder $J - K_S$ by $\sim$ 1.5 $\sigma$ for the stars with reported hot excesses, compared to those without.

Although our result is not sufficiently strong to constitute a completely independent confirmation of the existence of the 
hot excesses, it supports their reality.

\section{Hot dust emission properties and constraints}

In the following section, we develop general constraints that a successful model for the $K$-band excesses must meet. 
As discussed in the Introduction, we assume that they arise as thermal emission from hot dust, rather than through 
scattering or by some other mechanism. We then compare these constraints in detail with previous work based on conventional 
collisional cascade debris disk models. 

\subsection{Exploration of parameter space}
\label{sec:exploration}

The primary observational constraint on models for the hot dust emission is the slope of the spectral energy distribution 
(SED) of the hot emission between 2 and 10 $\mu$m. The slopes are not identical for all stars - for example, the very blue 
slope indicated for Vega appears to differ significantly from the redder slope for $\eta$ Crv\footnote{It is possible that 
the hot dust emission varies with time, complicating determining spectral slopes. In addition, the 10 $\mu$m measurements 
may have varying amounts of exozodi emission not related to the hot dust component}. However, three relatively well-measured 
systems, Vega, $\beta$ Leo, and 10 Tau, have excess slopes as blue as, or bluer than, the photospheric SEDs of their 
(relatively hot) stars. Since these cases will be the most challenging for models, we adopt the constraint that the 
emitting dust should have a SED roughly Rayleigh-Jeans in behavior.  

We used the Debris Disk Simulator (DDS) \citep{wolf2015} to run an array of simple and relatively unconstrained models to 
understand the permissible ranges of key parameters: 1.) dust sublimation temperature; 2.) dust composition; 3.) minimum 
grain size; 4.) grain size distribution; and 5.) structure of the system - whether the dust is in a ring or a broader disk. 
We considered three grain compositions: carbon, as the best known grain to survive in high temperatures, MgO, and 
FeO\footnote{We used the DDS default optical constants for carbon and FeO and obtained constants for MgO from 
\citet{palik1997}. A variety of grain materials of the type Mg$_{x}$Fe$_{1-x}$O have similar optical behavior to FeO.}, 
thermally robust compounds produced in the thermal breakdown of silicates or sulfides \citep{mann2007}. All models were 
run for an early A-star, T =  9000 K and L = 16 L$_\odot$. 

The most favorable model systems consisted of very small grains in a narrow ring around the star. For this geometry, 
we explored the effect of different assumed sublimation temperatures, assuming optical constants appropriate for carbon, 
nanograins between 5 and 6 nm in size, and in a ring extending from the sublimation radius to 1.5 times this radius. 
Sublimination temperatures of $\sim$ 1300 K or higher were sufficient to yield nearly Rayleigh-Jeans behavior between 2 
and 10 $\mu$m\footnote{For the rest of the paper, we adopt sublimation temperatures of 2000, 1800, and 1900 K respectively 
for C, FeO, and MgO \citep[e.g.,][]{lamoreaux1987, fedkin2006, kazenas2008}. As a check, we verified that these values 
gave sublimation distances consistent with those in \citet{mann2006}}. The primary effect of increasing sublimation 
temperatures was to decrease the mass in grains for a given output flux, by a factor of five from 1400 K to 2000 K and by 
another factor of two from 2000 K to 2600 K. At a sublimation temperature of 2000 K, a mass of about 10$^{15}$ kg in 5.5 nm 
grains was sufficient to yield a hot excess of about 1\% of the photospheric emission of an early A-star at 2 $\mu$m. 
Moreover, for small grains ($<$ 200 nm) we found that the SEDs are similar for C and FeO, dominated by the drop in emission 
efficiency at wavelengths significantly longer than the grain size\footnote{In this paper we quantify grain size in terms of 
radii.}. Pure MgO does not emit significantly at either 2 or 10 $\mu$m. We therefore concluded that these candidate materials 
in any combination (except pure MgO) could be responsible for the emission.

We next considered grain sizes and debris system geometries. In all cases we distributed the grains from the sublimation 
distance to an outer radius where there was a sharp cutoff in the grain density, and used the default density slope in the 
DDS of $-1.5$. We found that a system with a single grain size needed to have that size less than 200 nm to yield the 
Rayleigh-Jeans SED; similarly, power law (index $-3.5$) size distributions from 0.01 $\mu$m to 10 $\mu$m or larger values 
were not consistent with this SED. We also found that grains in a narrow ring worked much better than distributing them 
into a disk with an outer radius even as small as 1 AU. 

\subsection{Results from modeling specific hot excess systems}

Recent publications describing specific models of hot debris systems confirm the conclusions from this general exploration 
of parameter space \citep[e.g.,][]{absil2006,akeson2009, defrere2011, lebreton2013, vanlieshout2014}. \citet{absil2006} and \citet{defrere2011} showed that grain populations strongly 
emphasizing small sizes (a size distribution slope of $-5$ down to minimum grain sizes of 0.01 - 0.2 $\mu$m) at distances of 
predominantly 0.2 - 0.3 AU were needed to account for the hot excess around Vega. \citet{akeson2009} found that 
the most natural fits to their measurements of hot dust emission around $\beta$ Leo and $\zeta$ Lep required small 
(0.1 $\mu$m) refractory grains in narrow rings around the stars.  \citet{lebreton2013} found that the hot 
excess around Fomalhaut was best explained as due to a population of very small (size distribution slope of $-6$ from 0.01 
to 0.5 $\mu$m) grains presumably mostly of carbon, and originating in a narrow ring near 0.09 - 0.24 AU. In all of these 
models, it was difficult to retain the very small grains close to the star in the face of radiation pressure force. 

\citet{vanlieshout2014} tested whether the issue of grain blow out could be solved with a detailed model of grain inflow. 
They considered both analytic and numerical analyses, with inclusion of the tendency of the grains to pile up to high 
density at the sublimation radius as the grains are reduced in size and their $\beta_{\rm rad}$ (= photon pressure 
force/gravitational force) values increase, reducing the effective gravity they feel \citep{burns1979, kobayashi2009}. 
Their models had a parent body belt at a radius of 30 AU and with a particle size distribution of $n(s) \propto s^{-3.5}$ (where $s$ is the particle radius) 
undergoing a collisional cascade, and then followed (in the numerical case) the flow of the products inward to the 
sublimation radius, where the size distribution was distorted by wave patterns to favor grains just above the blow out 
size. The calculations, carried out in great detail, encountered two fundamental problems: 1.) the SEDs produced were far 
too red to match the Rayleigh-Jeans slope shown by some of the hot-dust sources; and 2.) even with inclusion of the grain 
pileup at the sublimation radius, the emission was more than an order of magnitude too faint to account for the observed 
excesses at 2 $\mu$m. 

\section{Hot excesses from magnetically trapped nanograins}

The above discussion illustrates the difficulties in accounting for the 2 $\mu$m excess emission with conventional debris disk theory.   
An alternative possibility is suggested by the demonstration by \citet{mann2006} that nanograins in the inner Solar System could acquire an 
electrical charge and be trapped by the solar magnetic field through Lorentz force.  In this section we show how a similar process could lead to the hot debris systems. 

Many of the hot excesses are around A-type stars, which are often thought not to have magnetic fields and hence should not generate Lorentz forces. 
However, the discovery of weak fields around Vega \citep{lignieres2009} and Sirius \citep{petit2011} showed this picture to be oversimplified. 
These fields are described by a ``failed fossil" hypothesis in which the field left from the formation of the star is still evolving towards equilibrium. 
Under this hypothesis, A and B stars in general would have magnetic fields, but with values dependent on the stellar age, rotation rate, and initial field \citep{braithwaite2013}. 

\subsection{Grain charging}

Small grains near hot stars will quickly acquire electrical charge either through the photoelectric effect, or by thermionic 
emission \citep{lefevre1975}. They will therefore inevitably interact with the stellar magnetic field, if there is one. 
The process of photoelectric grain charging has been modeled in detail by \citet{pedersen2011}; although their paper is 
focused on the effect around young stellar objects, the calculations for a 10,000 K blackbody are directly applicable to 
the situation for hot grains around main sequence A-stars. Although they assume grain properties for silicates, their 
results should be qualitatively applicable to the other grain materials under consideration for the hot emission, since 
the work functions for these materials are generally similar or smaller than those of silicates (see \citet{weingartner2001} 
for detailed comparisons of the behavior of silicates and carbonaceous material). Their numerical model accounts for the 
capture cross section for the incoming photon according to Mie theory, bases the electron yield on the prescription of 
\citet{weingartner2001}, and includes the possibility that the resulting electric field from the positive charge of the 
grain is sufficient to cause it to recapture electrons that have escaped through the photoelectric effect. They find 
(their Figure 3) that a 30 nm grain in the radiation field of a 10,000 K blackbody very quickly (less than a second) 
acquires a charge of more than 100 electrons. For a hotter radiation field, their model shows a limiting charge of 
$q = s(\rm nm)$$ \times 27.5 ~ e^-$ (where $q$ is the total charge on a grain, $s$ is the grain radius 
and  $e^-$ is the electronic charge), acquired in less than 0.1 second. 
Similar charge levels have been derived by \citet{ma2013} and a similar size dependence has been found by 
\citet{ignatov2009}. It is likely that similarly large charges will accumulate in a 10,000 K radiation field within a few seconds. 

\subsection{Forces affecting the nanograins}

We now construct a rough figure of merit to see when Lorentz force due to the stellar magnetic field is sufficiently 
strong to affect the dynamics of small grains. 
The dominant forces on small particles orbiting a hot star are gravitational, radiative, and 
electro-magnetic\footnote{Since winds are very small for typical early-type stars, their effects will 
be negligible.}. Particles can also be transported into the inner regions of circumstellar debris systems via PRD, another radiative effect, but it is a weak and slow-acting force. 
Because radiation pressure and gravitational forces are both proportional to $r^{-2}$, they
are treated together and characterized through their ratio, $\beta_{\rm rad}$ \citep{burns1979}. The combined force on a grain is then 
$1-\beta_{\rm {rad}}$ times equation (18) of \citet{burns1979}, i.e., 

\begin{equation}
F_{\rm{grav+phot}}=\left( 1-{\beta}_{\rm{rad} }
\right)\frac{4}{3}\frac{\pi  \thinspace s^{3} \rho \thinspace GM_{\ast }}{r^{2}}
\end{equation}

\noindent
where $\rho$ is the density of the grain and $s$ its radius, $M_{\ast}$ is the mass of the star, and $r$ is the 
distance of the grain from the star. The Lorentz force can be taken to be the vector product ${\mathbf v} \times {\mathbf B}$; we take the field strength $B$ to have the radial dependence of a 
dipole ($\propto r^{-3}$). It can then be shown that the ratio of gravitational and photon forces on the grain to the 
Lorentz force is

\begin{equation}
\frac{F_{\rm grav+phot}}{F_{\rm Lorentz}} \approx 1.4 \frac{\left(1-\beta_{\rm rad}\right)\left(
\frac{M_{\ast}}{M_{\odot}}\right)^{1/2}\left(\frac{s}{\rm nm}\right)^2
\left(\frac{r}{\rm AU}\right)^{3/2}}{k_q\left(\frac{v}{v_{Kepler}} \right) \left( \frac{B_{\ast}}{\rm Gauss}\right)
\left(\frac{R_{\ast}}{R_{\odot}}\right)^3}\;,
\end{equation}

\noindent
where $k_q$ is the 
coefficient of the charge on the grain ($q = k_q \times s$) taken to have a density of $\rho$ = 2.7  g cm$^{-3}$, $B_{\ast}$ is the magnetic field at the surface of the 
star, $R_{\ast}$ is the stellar radius, $v$ is the velocity of the field relative to the grain, and $v_{Kepler}$ is the Keplerian velocity in the absence 
of any Lorentz forces. The magnetic field will rotate with the star out to a
radius of $\sim$ 0.2 AU (Section \ref{sec:fieldstructure}). For most A-stars, the local 
velocity of the rotating field will exceed the Keplerian velocity at all relevant radii. 

\begin{figure}
\begin{center}  
\includegraphics[angle=0,scale=0.8]{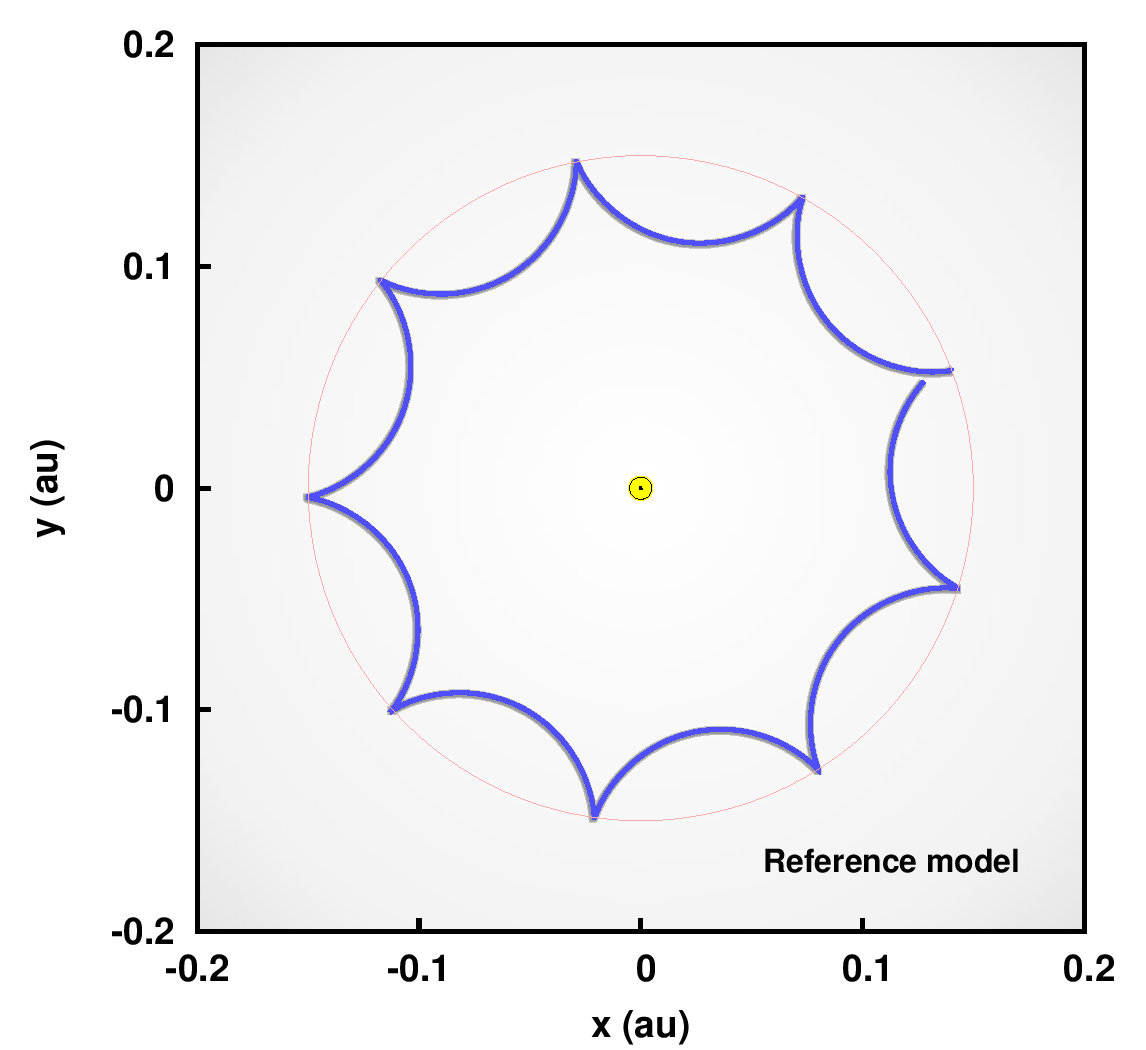}
\caption{The first circumstellar orbit of the nanograin in the Reference model is shown in blue, 
following its production at the sublimation point. The thin red line shows the sublimation radius. The gyroradius is 
half the distance from this line to the innermost part of the epicycles. The point of 
view is looking down on the epicyclic orbit from a perspective above the pole of the star. The parameters of 
the reference model are summarized in 
Table \ref{tab:reference}.}
\label{fig:ref_orbit}
\end{center}
\end{figure}

For a grain near the blow out size around an early A-star, e.g., 
$s$ = 5000 nm, at a distance of 0.2 AU and with a surface magnetic field of 1 Gauss, Equation (2) shows that the 
Lorentz forces are completely negligible, by a factor $> 10^{5}$. However, Equation (2) shows the strong 
dependence on the grain size, and for a 10 nm grain at the same position (and with $k_q \sim 27.5$), Equation (2) 
indicates that the Lorentz force is dominant. Consequently, when a nanograin 
is launched into interplanetary space around the star, it will quickly become charged and then will be picked up by 
the magnetic field and will orbit the star with the field while performing Lorentz-force-confined epicycles around 
the field lines (Figure \ref{fig:ref_orbit}): the nanograin is trapped against being blown out by radiation pressure 
force.

\subsection{Structure of stellar magnetic fields}
\label{sec:fieldstructure}

The magnetic fields around A-stars must be reasonably well-organized or they would not be detectable in polarimetry 
of the full visible stellar disk. Although they may have somewhat more complex geometry \citep{alina2012}, for simplicity 
we have assumed they are dipolar, i.e., fall off with distance, $r$, from the star $\propto r^{-3}$. 

By analogy with the Solar System, we expect the magnetic field around early-type stars 
to have characteristic domains. Near the 
stellar photosphere, the magnetic field will be governed by the radial motion of the stellar plasma, whose density 
decreases rapidly outward. The second domain is where the magnetic field energy density is greater than the plasma 
energy density, therefore controlling the angular rotation of the field. The magnetic field will therefore be rotating 
with the central star in this domain. Further out, in the third domain, the magnetic field energy density decreases to 
values smaller than the stellar wind energy density, resulting in the formation of so called ``Parker spirals'' 
\citep{parker1958,schatten1969}.

The nanograins in our models will either be in the second or third domain, as the first domain is very narrow and close 
to the photosphere. If they are in the third domain, we can assume the transverse velocity of the magnetic field to be equal 
to its value at the boundary between the second and third domains. The boundary of domains two and three can be found
by equating the magnetic field and plasma kinetic energy densities,
\begin{eqnarray}
\eta_{\rm magnetic} &=& \frac{B(r)^2}{2\mu_0}\;, \\
U                   &=& \frac{1}{2}\rho(r){v}_{\rm sw}(r)^2\;,
\end{eqnarray}
with
\begin{eqnarray}
B(r)    &=& B_{\ast}\frac{R_{\ast}^3}{r^3}\;,\\
\rho(r) &=& \frac{\dot{M}}{4 \pi r^2 {v}_{\rm sw}(r)}\;,
\end{eqnarray}
where $\eta_{\rm{magnetic}}$ is the magnetic field energy density, $U$ is the plasma kinetic energy density, 
$B(r)$ is the magnetic field strength at a distance $r$ from the star, decreasing in strength as a dipole, 
$\rho(r)$ is the stellar wind matter density as a function of distance from the star, 
$v_{\rm{sw}}$ is the stellar wind speed as a function of distance from the star, and 
$\mu_0$ is the vacuum permeability. The boundary between the regions can then be found at
\begin{equation}
r=\left[\frac{4\pi B_{\ast}^2 R_{\ast}^6}{\mu_0 \dot{M}{v}_{\rm sw}(r)}\right]^{\frac{1}{4}}\;.
\end{equation}
The exact behavior of the stellar wind or the mass loss rate of early-type stars is not known. 
Theoretical models \citep[e.g.,][]{babel1995} 
predict mass loss rates around 2 orders of magnitude less than the solar one, at 
$\dot{M}\approx10^{-16}~M_{\odot}~{\rm yr}^{-1}$. Using these values, we calculate the boundary between domains 2 and 3 
to be at 0.19 AU from an early-type star (reference model introduced in Section \ref{sec:numerics}), near the sublimation 
radius. In Section \ref{sec:refmodel}, we explore the effects of varying the stellar rotation rate, therefore effectively 
also exploring the effects of the boundary being located closer to the star.

\subsection{Gyroradius and maximum trapped grain size}
\label{sec:refmodel}

We now investigate the orbital stability of the nanograins produced near the sublimation radius. 
We focus on the particle size range that can be trapped
in the rotating stellar magnetic field and the timescale to spiral inward to the central star. For simplicity, we 
assume the nanograins originate at the sublimation radius from large particles that are dragged inward from an 
external location. The sublimation temperature is taken to be between 1300 and 2300 K, which corresponds to a 
narrow distance range of 0.2 - 0.07 AU from an A0 spectral-type star. Initially, the nanograins adopt the orbital 
velocity of the parent particle. The primary loss mechanism is collisions, for which the probability depends on the 
nanograin volume density. To estimate the collision rate, we will also model the spatial distribution of particles.

\begin{figure}
\begin{center}  
\includegraphics[angle=0,scale=0.68]{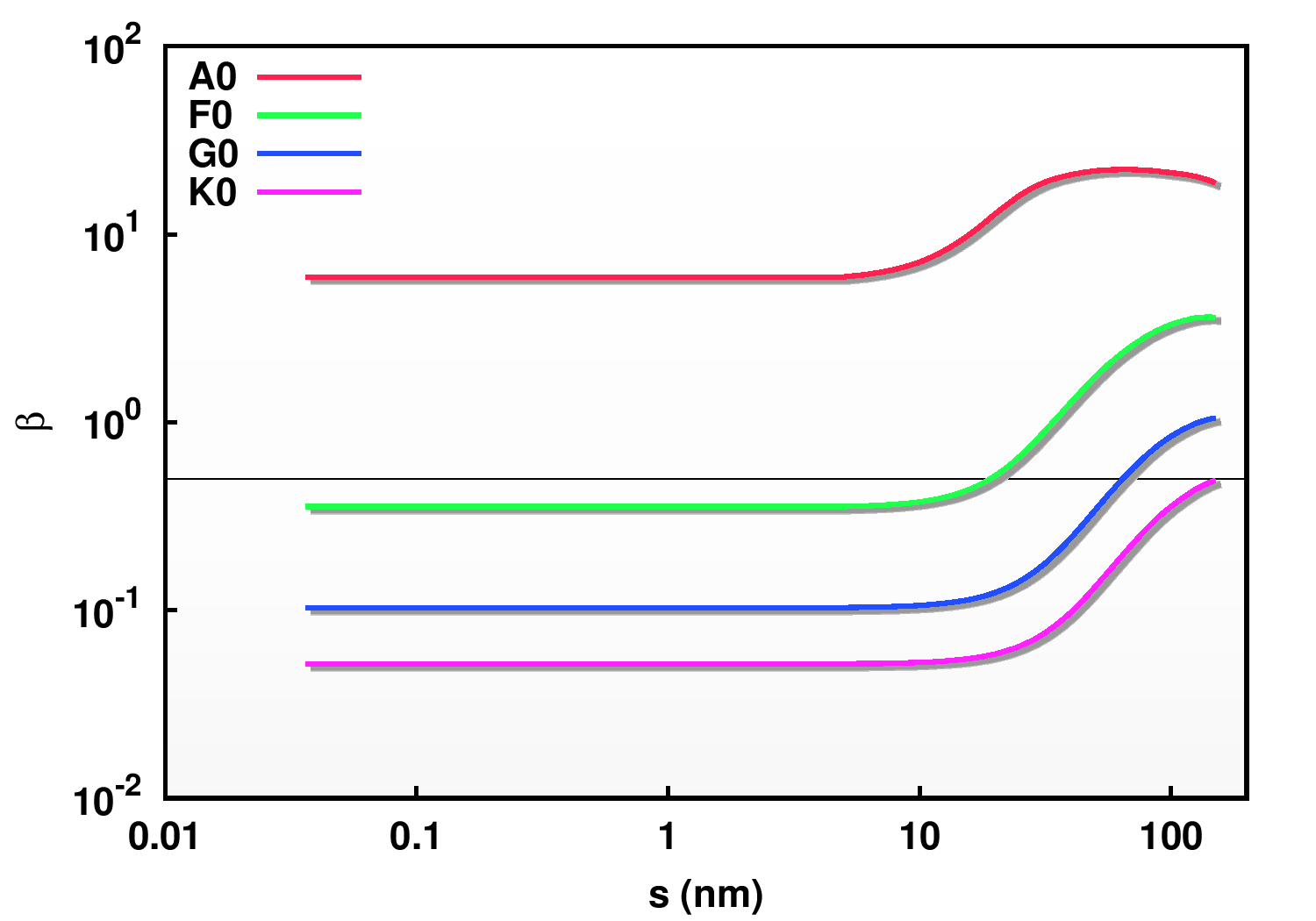}
\caption{The values of 
$\beta_{\rm rad}(s)$ used in section \ref{sec:deps} for various spectral-type sources.}
\label{fig:betas}
\end{center}
\end{figure}

We set a reference model roughly representative of most hot-excess systems, as summarized in Table \ref{tab:reference}. 
Since $\beta_{\rm rad}(s)$ levels off to a constant for all nanograins smaller than around 20 nm (see Figure \ref{fig:betas}), we assume it to be 
independent of grain size for all the models in this section, although will use size dependent values for models in 
Section \ref{sec:deps}.

\begin{deluxetable}{llr}
\tablewidth{240pt}
\tablecolumns{3}
\tabletypesize{\tiny}
\tablecaption{Parameters of the Reference model\label{tab:reference}.}
\tablehead{
\colhead{Variable} & \colhead{Description} & \colhead{Fiducial value}}
\startdata
$M$ 				&  stellar mass 				& 2.4 $M_{\odot}$	\\
$B_{\ast}$			&  magnetic field strength at stellar surface	& 1.0 Gauss		\\
$R_{\ast}$			&  stellar radius				& 2.3 $R_{\odot}$	\\
$r_{\rm sub}$			&  sublimation radius 				& 0.15 AU		\\
$s$				&  nanograin radius 				& 5 nm			\\
$\rho$				&  nanograin density				& 2.7 g cm$^{-3}$	\\
$\beta_{\rm rad; nano}$		&  nanograin radiative coefficient		& 6.0			\\
$\beta_{\rm rad; parent}$ 	&  parent particle radiative coefficient	& 0.2			\\
$P_{\rm rot}$			&  rotation period of star			& 10 hours		\\
 $\Omega_{\rm ast}$                                   & angular frequency of the central star  & $2 \pi P_{\rm rot}^{-1}$   \\
$k_{q}$				&  electric charge coefficient			& 27.5 e$^{-}$/nm
\enddata
\end{deluxetable}

The motion of the magnetically trapped nanograins is described to first order by their epicycles with amplitudes equal to the gyroradius 
(Larmor radius or cyclotron radius), circling the star at a distance determined by the grain release point. The gyroradius of a particle defines 
the volume of space within which a particle of its size is confined; therefore it also determines the collisional timescale of the particle. Hence, 
it is important to understand the range of values it can take. The gyroradius can be found by equating the sum of the 
forces acting on the nanograin (gravitational, electro-magnetic Lorentz, radiative blowout and centripetal forces) to 
zero. For an analytic estimate, we will assume that PRD is negligible compared to the other forces. Therefore we can 
write
\begin{equation}
\frac{\mu\left(\beta_{\rm rad;nano}-1\right)}{r_{\rm sub}^2} + \frac{q}{m}\left({v}_{\rm gyro} + {v}_{\rm orb} - \Omega_{\rm ast} r_{\rm sub}\right)B_{\ast}\frac{R_{\ast}^3}{r_{\rm sub}^3} + \frac{{v}_{\rm gyro}^2}{r_{\rm gyro}} + \frac{{v}_{\rm orb}^2}{r_{\rm sub}} = 0\;.
\end{equation}
The variable
$\mu$ is the standard gravitational parameter of the central star ($\mu = {\rm G} M_{\ast}$), $m$ is the particle mass, 
and $\Omega_{\rm ast}$ is the angular
frequency for the rotation of the central star. 
Here we determine the Lorentz and centripetal forces on the particle at its furthest location from the central star
on its epicycle. The charge a nanograin accumulates is $q = k_q s$. The orbital and gyro-rotational velocity a nanograin 
acquires depend on its inertia. Small nanograins will accelerate close to the speed of the magnetic field, while the 
largest trapped grains will remain near the parent particle's velocity:
\begin{eqnarray}
{v}_{\rm orb} ( s \ll s_{\rm max} ) &\approx& \Omega_{\rm ast} r_{\rm sub}\\
{v}_{\rm orb} ( s \sim s_{\rm max} ) &\approx& {v}_{\rm parent}\;,
\end{eqnarray}
where $s_{\rm max}$ is the largest trapped particle, $\Omega_{\rm ast} = 2 \pi P_{\rm ast}^{-1}$, and the orbital velocity of the 
parent particle is
\begin{equation}
{v}_{\rm parent} = \sqrt{\frac{\mu\left(1-\beta_{\rm rad; parent}\right)}{r_{\rm sub}}}\;.
\end{equation}
We approximate the gyrotational velocity of the small nanograins to be
\begin{equation}
{v}_{\rm gyro} ( s \ll s_{\rm max} ) \approx {v}_{\rm{parent}} - \Omega_{\rm ast} r_{\rm sub}\;.
\end{equation}
\noindent
i.e., assuming that the particle retains its initial velocity relative to the magnetic field, $v_{\rm{parent}}$ - $\Omega r_{\rm{sub}}$. 
The gyroradius of a small nanograin is then expressed by
\begin{equation}
r_{\rm gyro} = \frac{r_{\rm sub}^3 \left({v}_{\rm parent} - \Omega_{\rm ast} r_{\rm sub}\right)^2}{r_{\rm sub}\mu\left(1-\beta_{\rm r;n}\right) - \Omega^2 r_{\rm sub}^4 - B_{\ast}R_{\ast}^3\left({v}_{\rm parent} - \Omega r_{\rm sub}\right)\frac{q}{m}}\;.
\end{equation}

\noindent
The gyroradius of our reference model particle is 0.066 AU. The largest nanograin that the field can trap will remain in a circular orbit around the star with the velocity of the 
parent particle, since for this case the Lorentz force just cancels the radiative force. The centripetal force and gyrotational velocity are not defined in this case. 
This yields a largest particle size of
\begin{equation}
s_{\rm max}^2 = \frac{3 k_q B_{\ast} R_{\ast}^3\left(\Omega_{\rm ast} r_{\rm sub} - {v}_{\rm parent}\right)}{4\pi\rho\mu r_{\rm sub}\left(\beta_{\rm rad;nano}-\beta_{\rm rad;parent}\right)}\;.
\label{eq:smax}
\end{equation} 
Solving this equation for our reference model yields an upper limit of 121 nm on the size of the nanograins
trapped by the magnetic field of the reference system. A narrow range of larger particles, that would otherwise
be removed via radiative pressure blowout (if $\beta_{\rm rad;nano} > 0.5$), remain in the system with ``inverse'' 
epicycles outside of the sublimation point. The lower limit on the distribution is set by the molecular limit (a grain must consist of a number
of molecules) to be several tenths of a nm. 

\subsection{Numerical analysis of the orbital dynamics}
\label{sec:numerics}

We now augment and illustrate these arguments with a numerical model. 
The accelerations of a grain are:
\begin{eqnarray}
{\mathbf a}_{\rm gr} &=& - \mu \frac{\mathbf r}{r^3}\\
{\mathbf a}_{\rm rad~blow} &=& \mu \beta_{\rm rad} \frac{\mathbf r}{r^3}\\
{\mathbf a}_{\rm rad~drag} &=& -\frac{\mu \beta_{\rm rad}}{{\rm c} r^2}\left(\dot{r}\frac{\mathbf r}{r} + {\mathbf v}\right)\\
{\mathbf a}_{\rm em} &=& \frac{q}{m}\left({\mathbf v} - {\bf \Omega_{\rm ast}} \times {\mathbf r}\right)\times {\mathbf B}\;.
\end{eqnarray}
The various accelerations are: gravitational (${\mathbf a}_{\rm gr}$), radiative blowout (${\mathbf a}_{\rm rad~blow}$),
radiative drag - PRD (${\mathbf a}_{\rm rad~drag}$), and electromagnetic Lorentz (${\mathbf a}_{\rm em}$).  c is the speed of light and ${\mathbf v}$ is the relative motion of the particle w.r.t.\ the central star. 

By necessity, the analytic analysis neglected several important physical effects, an issue we address here. 
The trapped particles will experience minor perturbations from PRD, varying 
electro-magnetic forces along their gyro-orbits (due to the strong radial dependence of the 
magnetic field), and also varying gravitational force along the gyro-orbits. Furthermore, the orbital and
gyro-velocity of the largest trapped particles depend on the mass of the nanograins. We 
evaluate these effects with numerical methods, finding that they result in a slow orbital decay of the particles 
and also lead to smaller gyroradii for the larger nanograins still trapped by the magnetic field.

We initiate the numerical analysis by tracing the orbital evolution of a ``large'' ($\sim 10~\micron$) 
parent particle subject to the accelerations in Equations (15 - 18), which is dragged in via PRD from an external 
location without undergoing collisions. The exact initial orbit of the parent particle is irrelevant, as its orbit 
will be circularized by the PRD by the time it approaches the point of sublimation. At that point, we replace 
the parent particle with a charged nanograin, and thereafter its orbit will be mostly determined by the radiative 
blowout force and the electro-magnetic Lorentz force from its interaction with the rotating stellar magnetic 
field. These forces place the trapped nanograin on an epicyclic orbit.

We will assume the particles have an orbital inclination of $\iota = 0$ and that the stellar magnetic field 
vector is perpendicular to the particle orbital plane. We set the stellar magnetic field to be coupled to the 
stellar rotation (as described in section \ref{sec:fieldstructure}). The 
dynamical evolution of the particles is evolved using an adaptive step-size 4th order Runge-Kutta integrator,
with an allowed relative numerical error of $10^{-12}$ per step.

In Figure \ref{fig:ref_orbit}, we show the orbit of the 5 nm particle in our Reference model. The 
particle has a gyroradius of \mbox{$\sim 0.018$ AU}, with a gyro-period of 1.6 hours and an orbital period around
the star of 0.59 days. Therefore, the particle has roughly 9 epicycles in a single orbit.
The particle is orbiting in its epicycle at 340 km s$^{-1}$ and at 2773 km s$^{-1}$ around the star.
The gyroradius given by the numerical solution is far from the value given by the analytic approximation,
which yields 0.066 AU. As noted before, the offset is due to the approximation that the gyro-velocity will
be equal to the difference between the orbital velocity of the parent particle and the projected 
velocity of the magnetic field. 

\begin{figure}
\begin{center}  
\includegraphics[angle=0,scale=0.68]{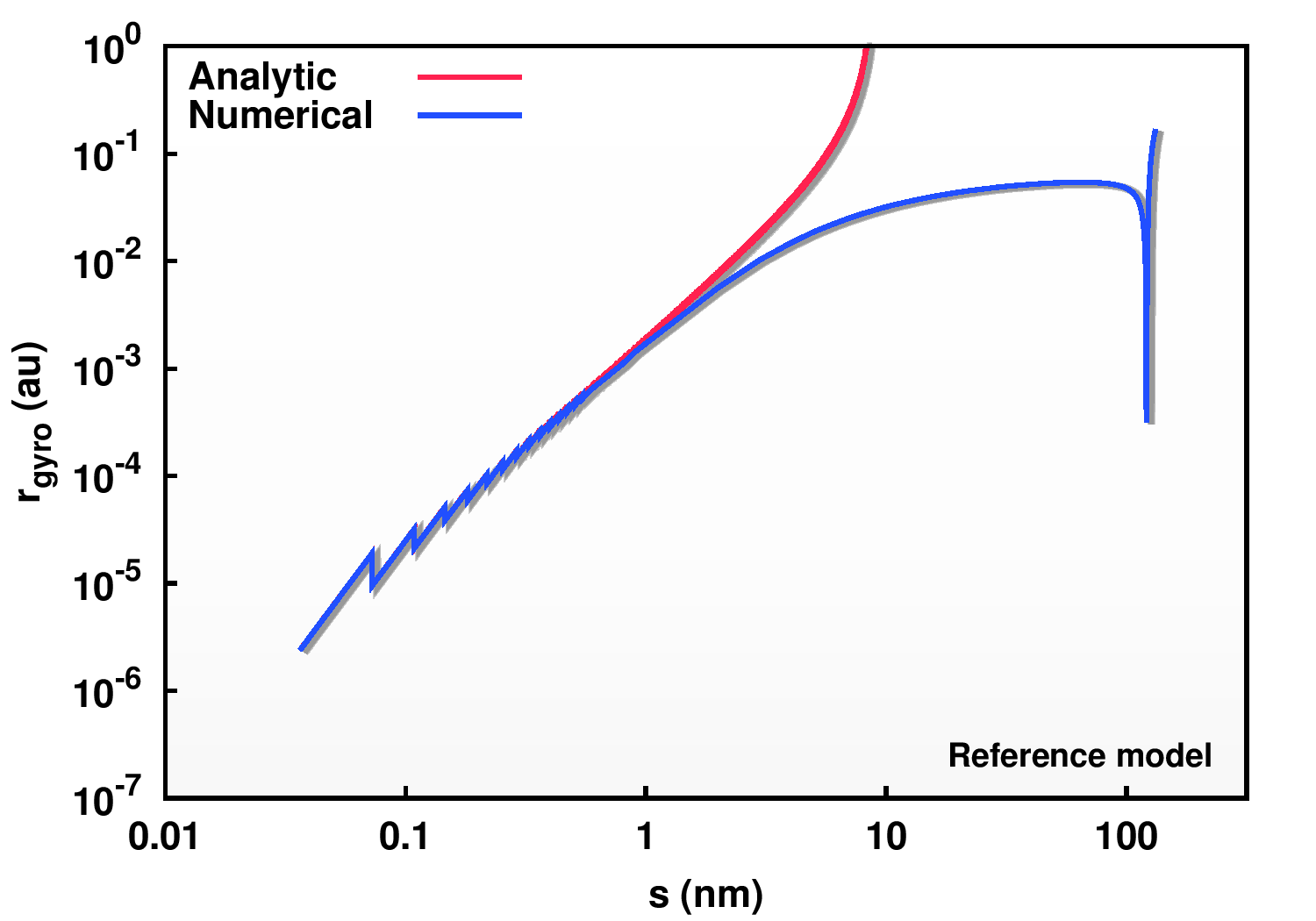}
\caption{The analytic and numerical solutions of the gyroradius as a function of particle size in the 
reference model, assuming $\beta_{\rm rad;nano}$ is independent of particle size. The ``sawtooth'' at 
the smallest sizes is due to the quantized nature of charge. The analytic estimate breaks down for larger
sizes, where the inertia of the particles is too large for them to be accelerated to the angular rotation
of the magnetic field. Furthermore, larger nanograins are also trapped in gyrotational orbits outside 
the sublimation point (see the break in the numerical analysis).}
\label{fig:gyro}
\end{center}
\end{figure}

\begin{figure}
\begin{center}  
\includegraphics[angle=0,scale=0.68]{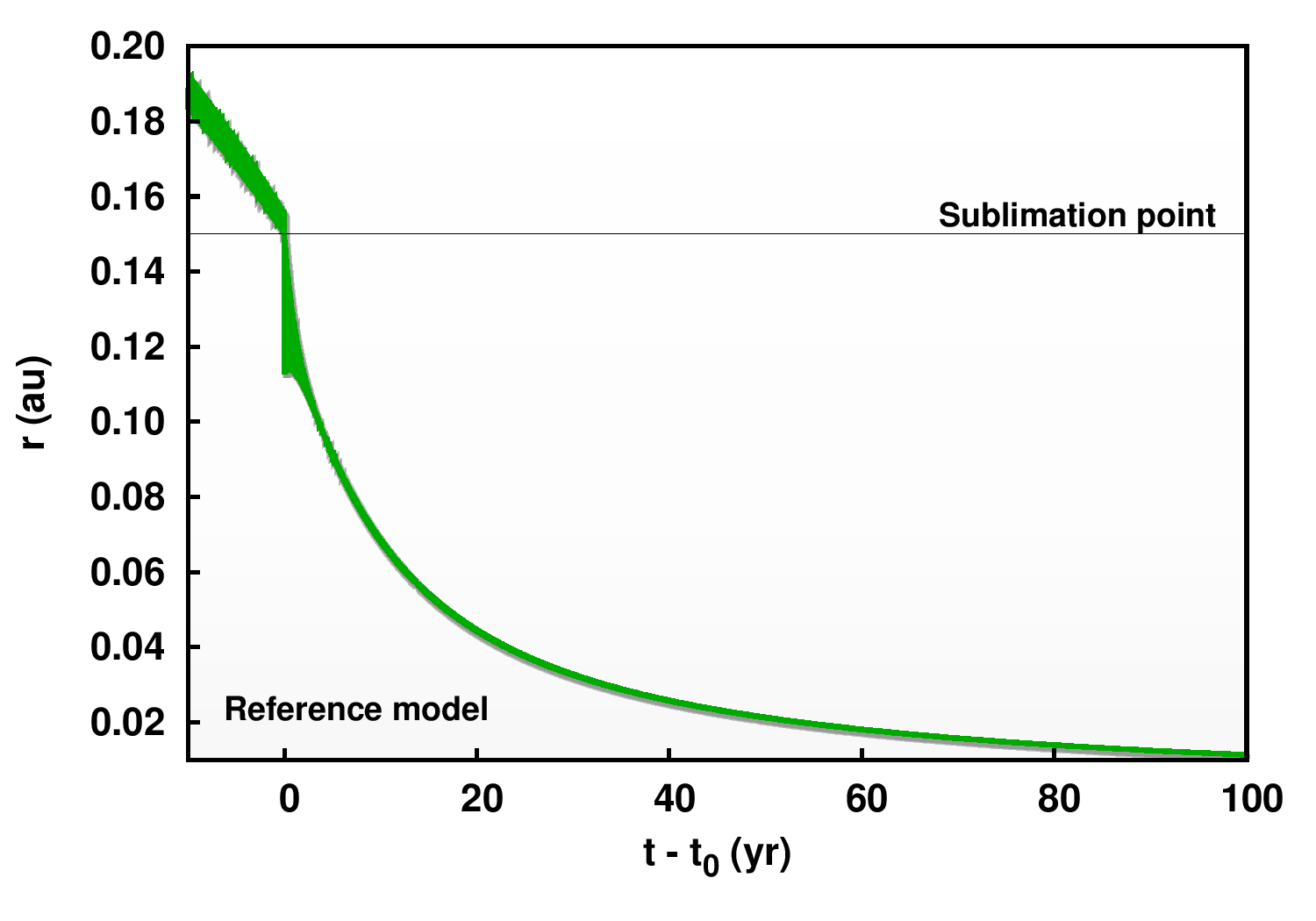}
\caption{The orbital decay of the nanograin in the Reference model. The radial distance before
sublimation varies due to the initial ellipticity (e=0.01) of the parent particle's orbit.}
\label{fig:ref_decay}
\end{center}
\end{figure}

In Figure \ref{fig:ref_decay}, we show the orbital decay of the particle, assuming it to be undisturbed 
by collisions. The nanograin of our reference model would remain in the system for a few decades. Finally, 
we plot both the analytic estimate and the numerical solution of the gyroradius as a function of particle 
size in the reference system in Figure \ref{fig:gyro}. The analytic approximation agrees with the numerical
solution up to a few nm in size, from which point the analytic solution yields larger gyroradii. 
The largest particle size given by the analytic analysis (Equation \ref{eq:smax}) agrees with the result from the
numerical model. The numerical analysis shows that particles larger than $s_{\rm max}$ are also placed on 
gyro-orbits outside of the sublimation point.

\section{Model dependence on variables}
\label{sec:deps}

In this section we investigate how the dynamical evolution of the nanograins depends on the system
variables. We are mainly interested in the size-range and location of the trapped particles and
their absolute velocities within the system to allow estimates of collisional timescales. We consider the 
effects of (1) the spectral-type of the central star; (2) the magnetic field strength; and (3) the rotation 
period of the central star. We do not investigate the dependence on the electric charge coefficient, 
$k_q$, as it is a multiplicative constant of the magnetic force, so varying it is equivalent to changing 
the magnetic field strength. Unlike in the previous section, 
we will assume $\beta_{\rm rad}(s)$ to vary with particle size to provide a more realistic 
understanding of the nanograin dynamics in various systems. We plot the $\beta_{\rm rad}(s)$
values we use in our models as a function of particle size and stellar spectral type in Figure
\ref{fig:betas}.

\subsection{Dependence on the spectral-type}
\label{sec:sptype}

We analyze the gyro-orbits of nanograins around A0, F0, G0, and K0 spectral-type stars in this subsection. 
The variables of the modeled systems are summarized in Table \ref{tab:sptype}. As Figure \ref{fig:betas} shows,
the $\beta_{\rm rad}(s)$ values of nanograins around most stars do not reach 0.5, i.e.\ the nanograins will not be 
blown out of the system by radiation pressure force. However, the interactions with stellar winds around the later 
type stars have a strong influence on the grains, creating a pseudo-PRD drawing grains inward and also carrying 
small grains out of the system. Grain charging by photoelectric effect is less effective for late-type stars, but 
charging can also occur through thermionic emission and electron impacts from the stellar wind plasma. Unless the 
grains are very hot, their charge is set by an equilibrium between the first (which drives their charge positive) 
and the second (which drives them negative). A detailed analysis for the Solar System is provided by \citet{mann2014}, 
where it is demonstrated that the resulting electrical charges can cause nanograins within $\sim$ 0.15 AU of the Sun 
to become trapped in the solar magnetic field against being carried away by the solar wind. The behavior around other 
stars of roughly solar type should be similar.

\begin{deluxetable}{lrrrr}
\tablewidth{240pt}
\tablecolumns{5}
\tabletypesize{\tiny}
\tablecaption{Variables of various spectral-type systems analyzed in section \ref{sec:sptype}.
The remaining variables are kept at the same values as in the reference model and the $\beta_{\rm rad;nano}$
is a function of particle size, as described in the text. Due to the size dependent $\beta_{\rm rad;nano}$,
$s_{\rm max}$ will take a different value from the analytic prediction.
\label{tab:sptype}}
\tablehead{
\colhead{Spectral type} & \colhead{$M_{\ast}$}    & \colhead{$R_{\ast}$}    & \colhead{$r_{\rm sub}$} & \colhead{$s_{\rm max}$} \\
\colhead{             } & \colhead{($M_{\odot}$)} & \colhead{($R_{\odot}$)} & \colhead{(AU)}          & \colhead{(nm)}}
\startdata
A0 [Reference model with $\beta_{\rm rad}(s)$] & 2.40	  & 2.30		    & 0.15		       & 62	\\
F0				& 1.70			  & 1.30		    & 0.13		       & 87	\\
G0				& 1.10		  	  & 1.05		    & 0.06		       & 141	\\
K0				& 0.78			  & 0.85		    & 0.03		       & 186
\enddata
\end{deluxetable}

\begin{figure}
\begin{center}  
\includegraphics[angle=0,scale=0.68]{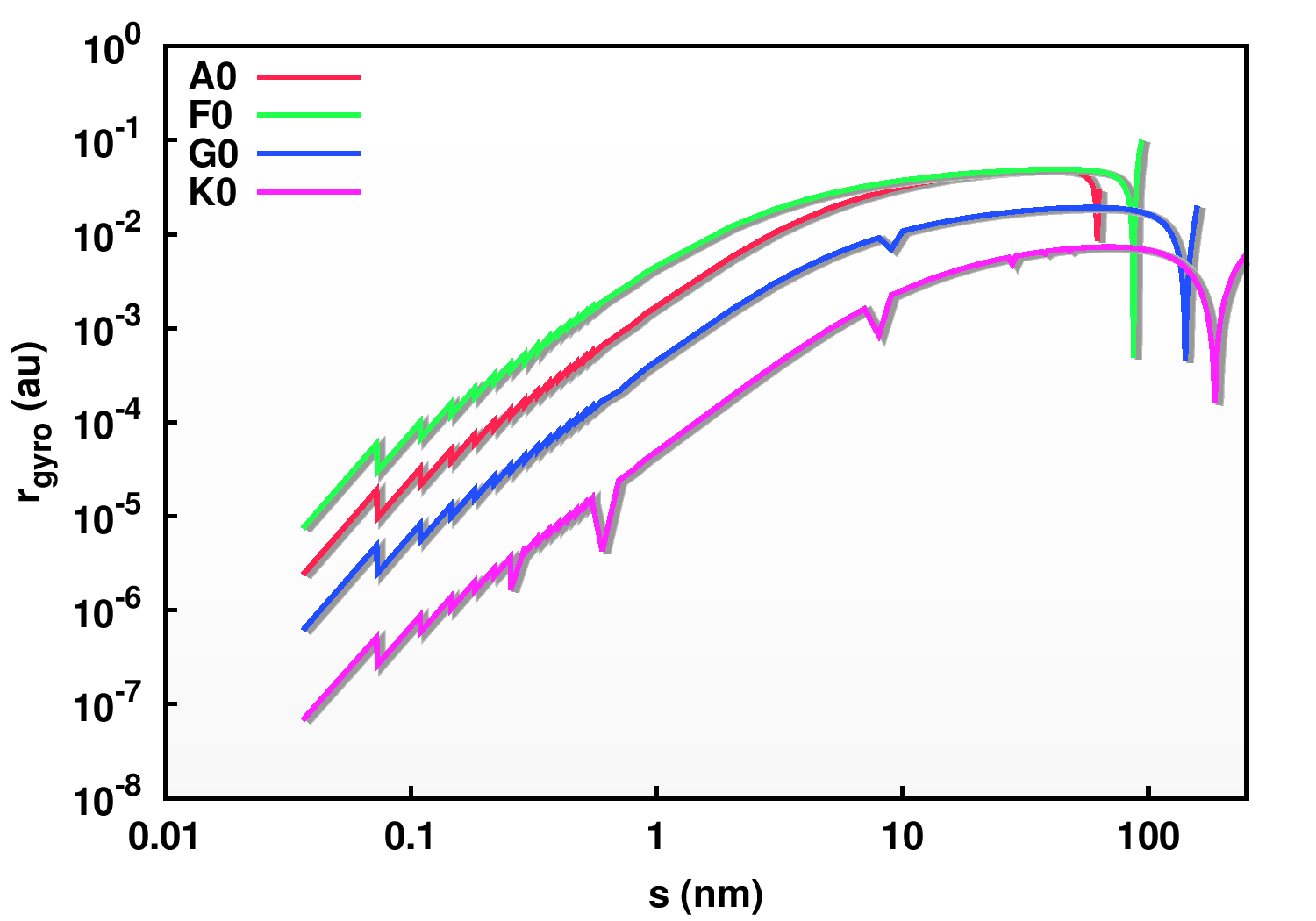}
\caption{The gyroradius as a function of particle size around various spectral type stars; in this case, we have used values for $\beta_{\rm {rad}}$ illustrated in Figure 3.}
\label{fig:sptypes}
\end{center}
\end{figure}

\begin{figure}
\begin{center}  
\includegraphics[angle=0,scale=0.68]{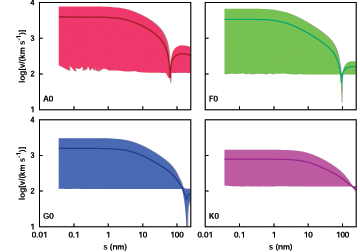}
\caption{The absolute velocity range as a function of particle size around various spectral type stars. The shaded 
region shows the range of velocities, while the line is the orbit-averaged value.}
\label{fig:sptypes_vel}
\end{center}
\end{figure}

Although the expected behavior is complex because of interactions with the stellar wind \citep{mann2014}, 
it is illustrative to examine the simple version provided by our model. In Figure \ref{fig:sptypes}, we plot the 
gyroradii of the particles in these systems. The largest particles the systems are able to retain increase with 
later-type stars. Furthermore, the characteristic collisional velocities decrease for later-type stars, as shown in 
Figure \ref{fig:sptypes_vel}. Therefore, the effects of magnetic field trapping could be more evident around them, as 
the collisional probabilities of the particles decrease with the collisional velocities. It is worth noting though, that
the models presented here maintain the fast 10 hour rotation of the Reference model which may not be a common value 
for most later-type systems. Increasing the rotation period will decrease the effects of magnetic field trapping and 
thus the production of hot excess (see Section 5.3).

\subsection{Dependence on the magnetic field strength}
\label{sec:fieldstrength}

\begin{figure}
\begin{center}  
\includegraphics[angle=0,scale=0.68]{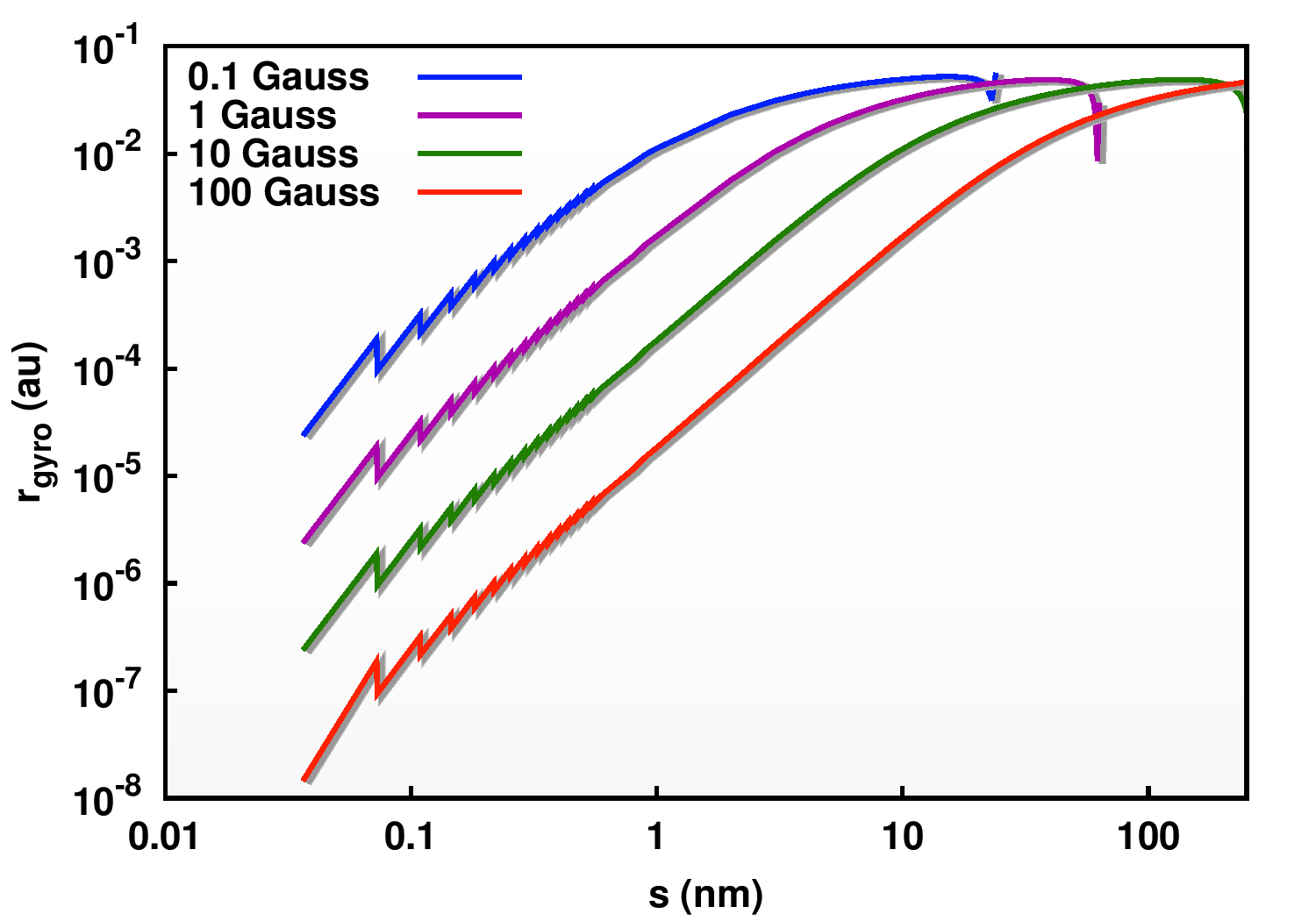}
\caption{The gyroradius as a function of the magnetic field strength for various size particles in the reference model, 
with $\beta_{\rm{rad}}$ as in Figure 5.}
\label{fig:Bfields}
\end{center}
\end{figure}

\begin{figure}
\begin{center}  
\includegraphics[angle=0,scale=0.68]{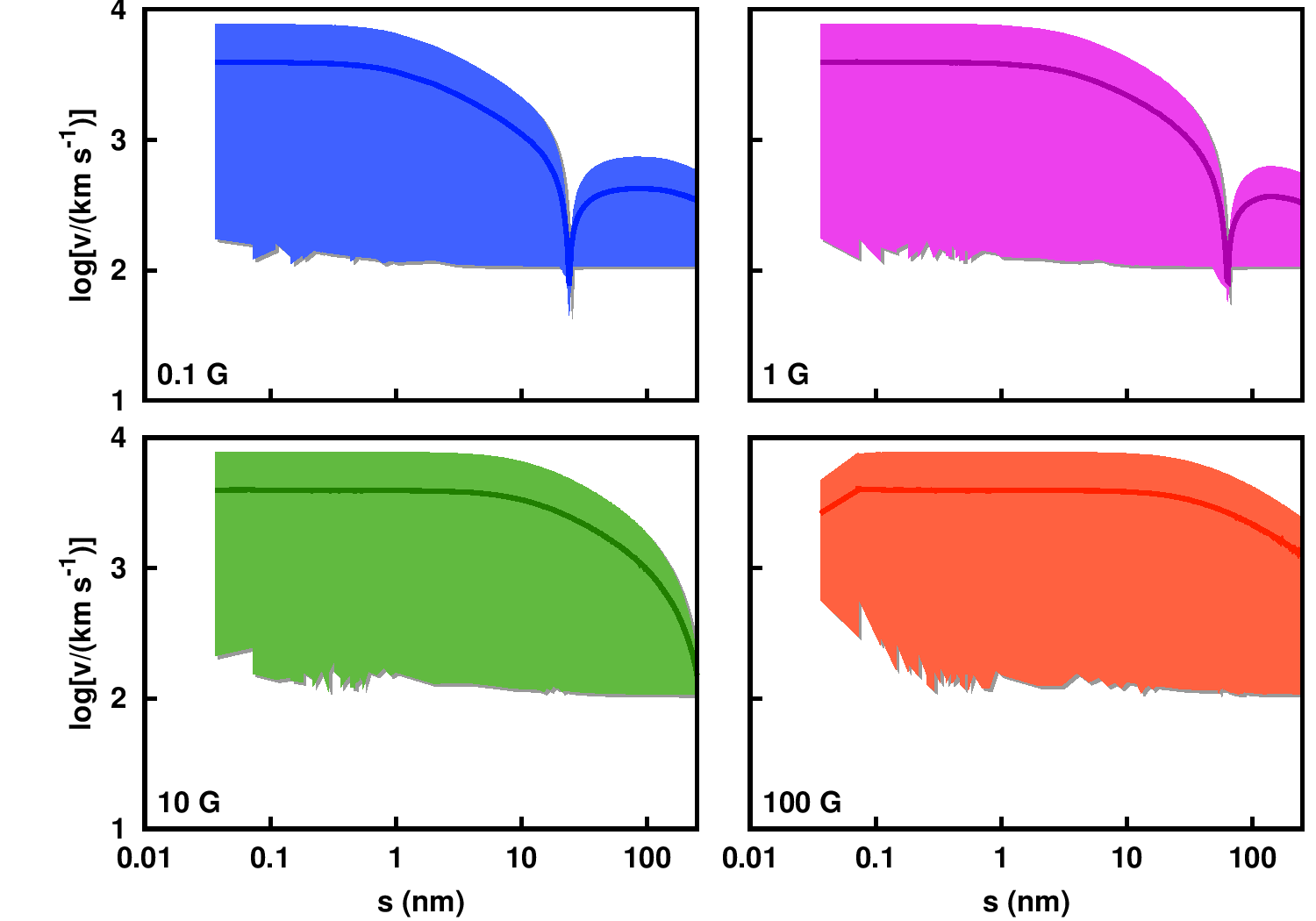}
\caption{The absolute velocity range (shaded region) as a function of the magnetic field strength for various size particles 
in the reference model. The lines show the orbit-averaged values}
\label{fig:Bfields_vel}
\end{center}
\end{figure}

In this subsection, we investigate the sensitivity of the trapping mechanism to varying magnetic field 
strengths. Peculiar Ap spectral-type stars are known to have over 100 G magnetic fields. Therefore, we model 
a three order of magnitude range of magnetic fields to study their effects, from 0.1 to 100 Gauss.
In Figure \ref{fig:Bfields}, we show the gyoradii of the particles around the A0 star (reference) model,
with size dependent $\beta_{\rm rad}(s)$ values.
The model allows large variations in the magnetic fields. A magnetic field even 
as weak as 0.1 Gauss\footnote{and therefore smaller than the fields of $0.2 \pm 0.1$ G measured for Sirius \citep{petit2011} and 
$0.6 \pm 0.2$ G for Vega \citep{lignieres2009}} is able to retain a relatively large range ($s$ up to 23 nm) of particle sizes 
near the sublimation radius. The velocities of the particles do not depend on
the strength of the magnetic field (Figure \ref{fig:Bfields_vel}), therefore the collisions around systems 
with strong fields will not be more violent than in systems with weak fields. Systems with stronger 
fields will be able to keep larger particles within their systems. However, the gyro-radii of the particles will be 
orders of magnitude smaller, resulting in much larger collisional probabilities and more rapid loss of the grains. 
The relative number of large particles is likely to be small (the size distribution from nanoparticle 
production mechanisms is thought to favor tens of nm) and hence the emission from a system where only 
the few large particles survive is likely to be weak.
Therefore, it is unlikely that large magnetic fields will lead systematically to strong hot excesses.

\begin{figure}
\begin{center}  
\includegraphics[angle=0,scale=0.68]{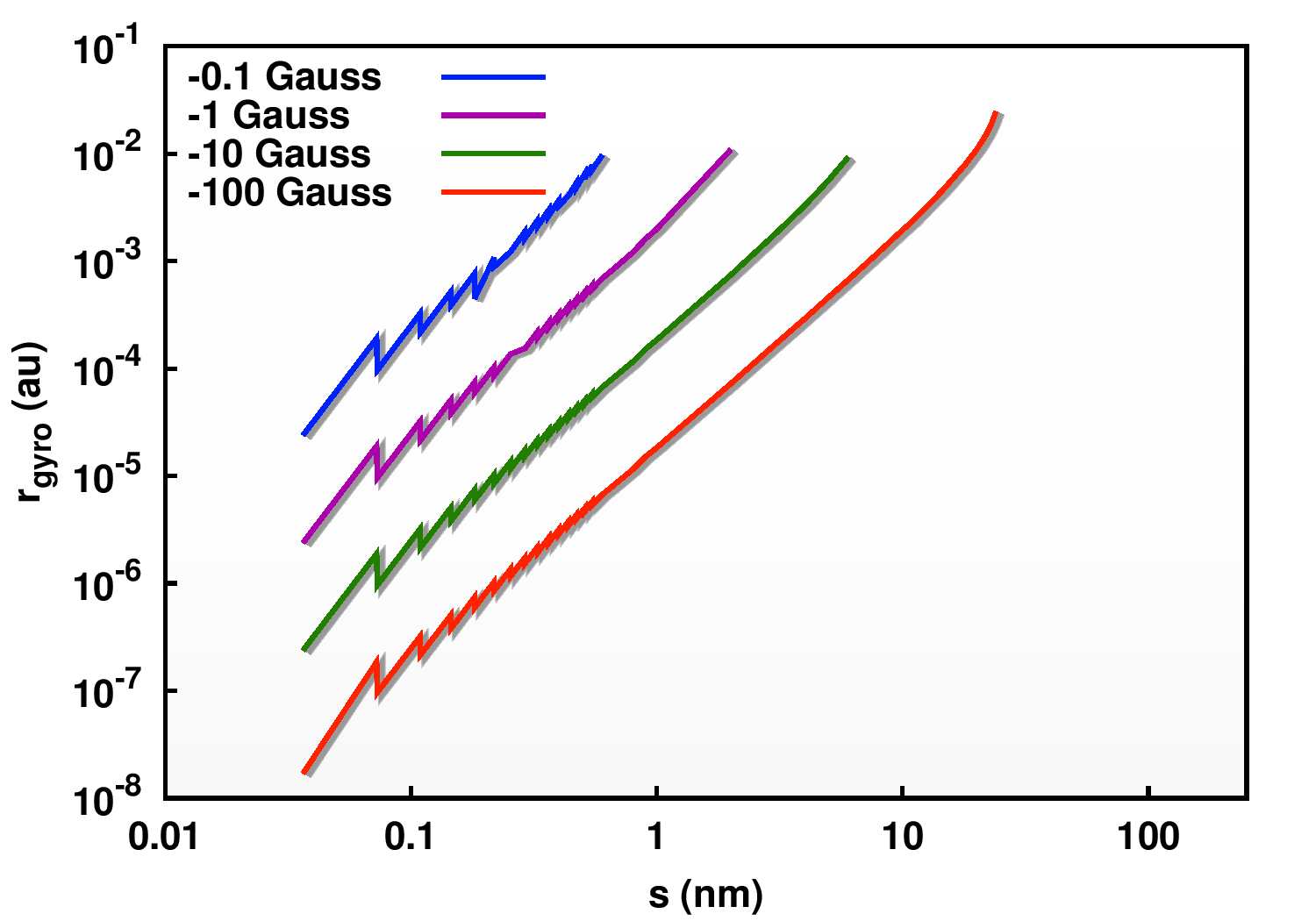}
\caption{The gyroradius as a function of the magnetic field strength for various size particles in the reference model, 
with the direction of the magnetic field flipped.}
\label{fig:neg_Bfields}
\end{center}
\end{figure}

The magnetic fields in early-type stars are primordial, therefore we do not have to model the
effects of periodic magnetic field flipping. However, systems where the rotational direction of 
the magnetic field and that of the nanograins are not aligned are plausible. In Figure \ref{fig:neg_Bfields}, we
plot the gyroradii of particles shown in Figure \ref{fig:Bfields}, but with the direction of the
magnetic field flipped. The system is still able to maintain particles within the system, although further
out from the sublimation radius on ``inverse'' epicycles and less effectively (over a smaller range of grain size).

\subsection{Dependence on the stellar rotation period}
\label{sec:rotation}

\begin{figure}
\begin{center}  
\includegraphics[angle=0,scale=0.68]{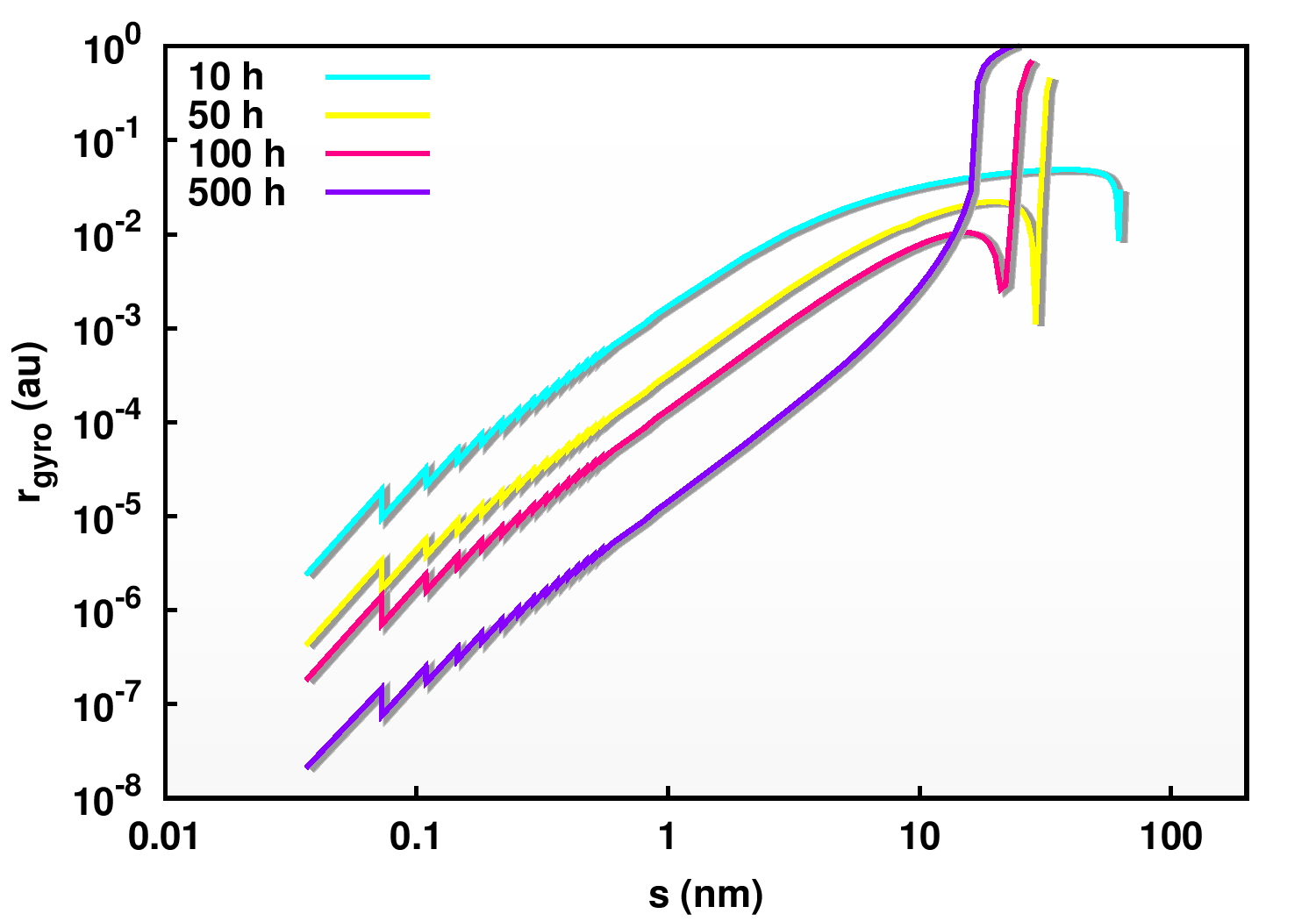}
\caption{The gyroradius as a function of the stellar rotation period for various size particles in the reference model.}
\label{fig:Trot}
\end{center}
\end{figure}

The faster a star rotates, the faster the magnetic field rotates with it and hence the stronger will be the 
electro-magnetic forces. In Figure \ref{fig:Trot}, we show the gyroradii of particles in the reference system, 
varying the rotation period of the central star. Within the typical rotation rate of early-type stars, nanograins 
of sizes up to 20-50 nm are expected to be trapped, with high velocities of motion, up to a few thousand km s$^{-1}$.

\begin{figure}
\begin{center}  
\includegraphics[angle=0,scale=0.68]{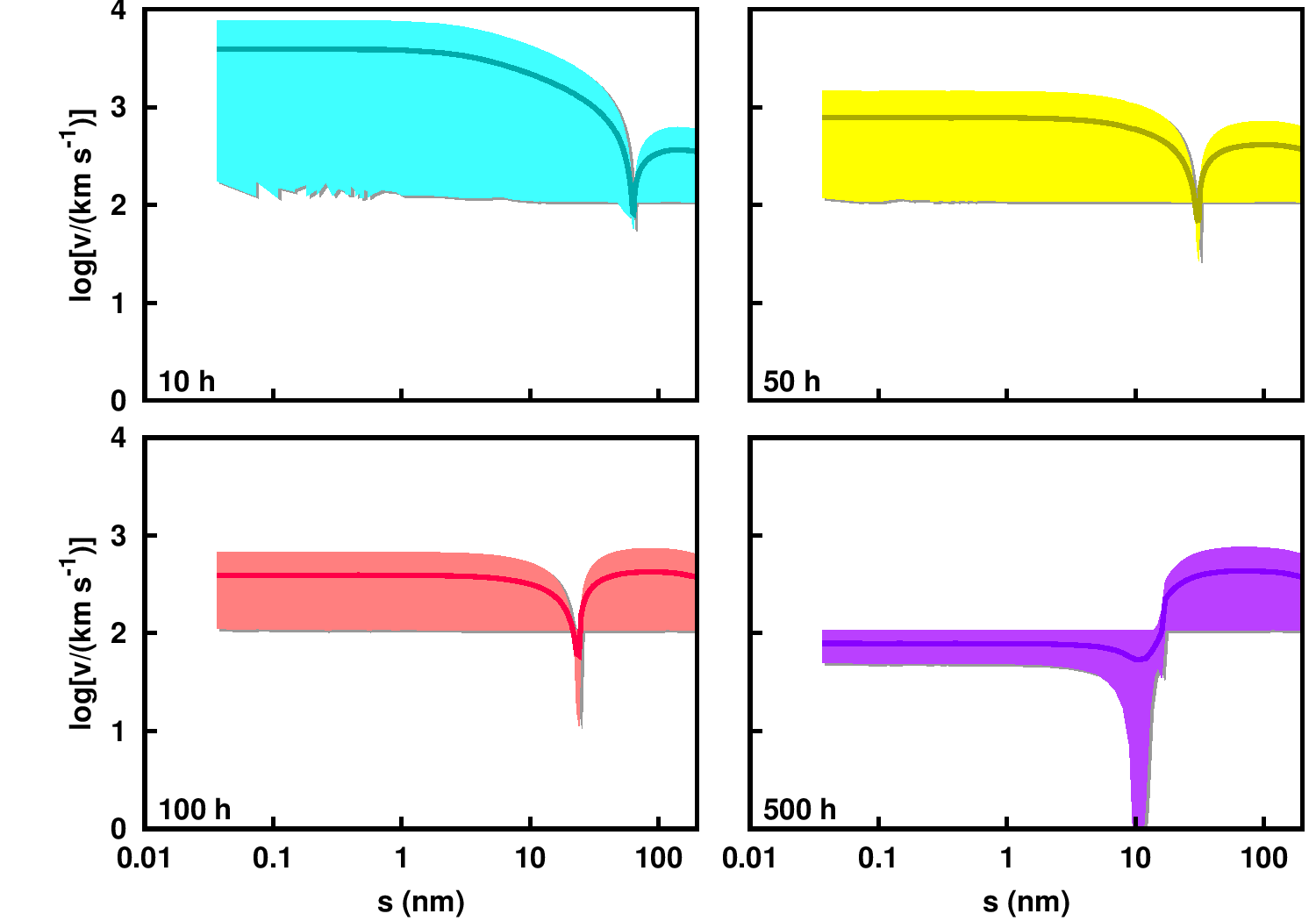}
\caption{The absolute velocity range (shaded) as a function of the stellar rotation period for various size particles in the 
reference model. The lines are orbit-averaged values.}
\label{fig:Trot_vel}
\end{center}
\end{figure}

Figure \ref{fig:Trot} shows that more rapidly rotating stars are able to trap larger nanograins. In Sections 6.2 and 
6.3 we show that these larger particles have smaller collisional cross sections and longer lifetimes; while their 
common orbital velocities will be large, they will interact at their gyroradial velocities, which will be smaller 
(due to the larger gyroradii for larger particles as shown in Figure \ref{fig:Trot}). The fast rotators therefore 
impose the minimum requirements on the mass transport to sustain the hot excesses. As a result, they should be most 
likely to show hot excesses.

\section{Mass Transport to the inner regions and the collisional timescale of nanograins}
\label{sec:steady}

The viability of our model depends on the total mass that is transported into and maintained 
at the inner regions near the sublimation radius. The relation between various timescales in 
the system determines the amount of dust continuously present in the inner regions. Below we 
give analytic estimates of the mass transport and collisional timescales of the nanograins,
and the steady-state levels of dust.

\subsection{Mass transport into the inner regions and nanograin production rate}
\label{sec:tprd}

To estimate the total amount of dust that can be theoretically transported into the 
inner regions of early-type star disks from outer asteroid-belt analog dust rings via PRD, we use 
the derivation in \citet{vanlieshout2014}, their Equation (11) (which assumes a single grain size). This 
yields $2\times10^{14}$ kg yr$^{-1}$ for our reference model from an outer region at 10 AU.

\subsection{Collisional timescale for nanograins}
\label{sec:times}

There are two significant loss mechanisms for nanograins in this situation: sublimation or sputtering, 
and grain-grain collisions. Sputtering will be modest around A-type stars given their small winds, but 
will be dominant around later types \citep{wurz2012}. Sublimation is not well understood 
because the grain properties may differ from those of bulk material of the same composition, and the 
composition of the nanograins is not well determined. We treated sublimation as in \citet{kobayashi2011}, 
derived for larger grains. 

The dynamical models presented in Section \ref{sec:numerics} show that the nanograins produced 
via sublimation are able to remain in the system for a considerable amount of time. The largest
nanograins de-orbit into the central star within a few decades. Since their production rate can be of 
order 10$^{14}$ kg yr$^{-1}$, compared with our rough estimate of a mass of 10$^{15}$ kg needed to produce a
typical hot excess (Section \ref{sec:exploration}), magnetic field trapping is a viable model to explain
this phenomenon if the nanograins are not destroyed via other mechanisms. The dominant loss 
mechanism for the nanograins is collisions, which we must quantify to understand the equilibrium grain density. 

The effects of collisions have been studied numerically by \citet{ohnishi2008}. They modeled collisions of grains 
of radii 1.4 and 4 nm with relative velocities of 3.6 - 6.1 km s$^{-1}$, and at temperatures appropriate for the 
interstellar medium. They found a variety of outcomes such as grain fusion, melting and nucleation. In general, 
grazing collisions did not destroy the grains, although as expected, direct ones did. A limited study of hydrocarbon 
nanograin collisions at higher velocities \citep{papoular2004} reached qualitatively similar conclusions. The mutual 
cross section for the collision of two identical grains is 4$\pi s^2$; however, to allow for survival of a significant 
fraction of a grain in a grazing collision, in the following we will assume that the cross section for grain destruction 
is the projected area of a single grain, $\pi s^2$.

\begin{figure*}[!t]
\begin{center}  
\includegraphics[angle=0,scale=0.68]{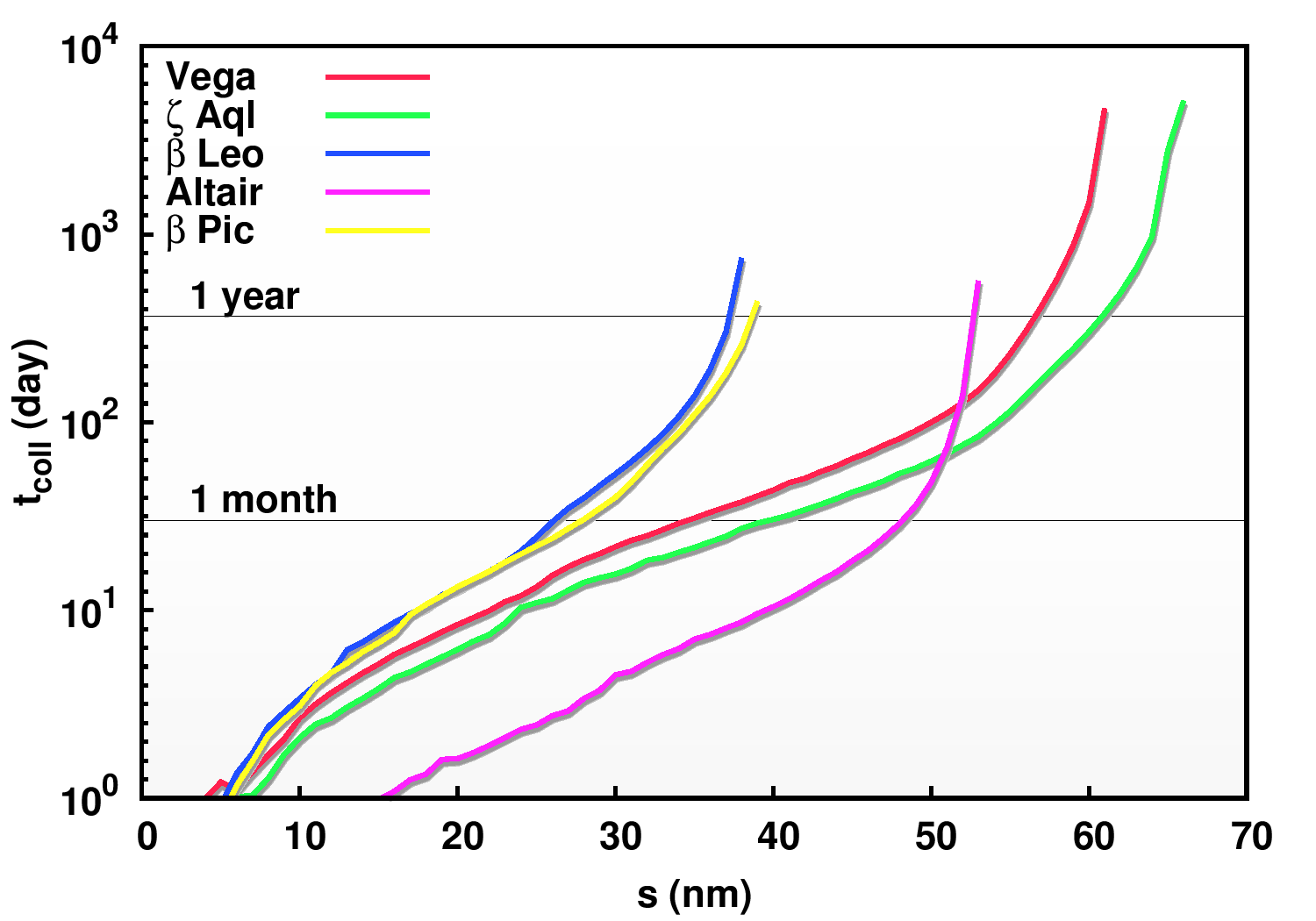}
\includegraphics[angle=0,scale=0.68]{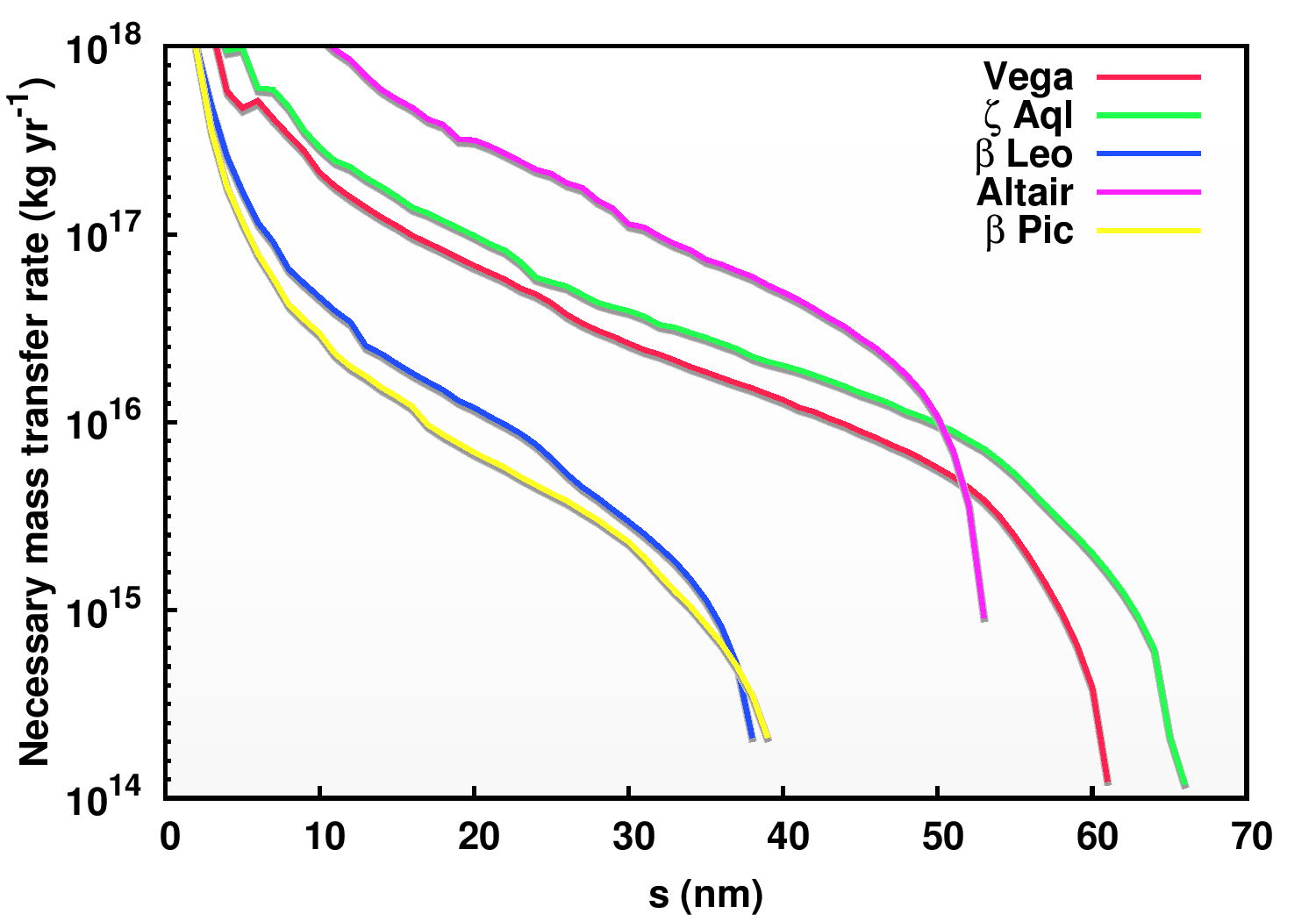}
\caption{The collisional timescale for some well known hot-excess systems assuming a single characteristic nanograin size
and the necessary mass transfer rate to allow a steady-state hot emission.}
\label{fig:tcoll}
\end{center}
\end{figure*}

The collisional timescale of a particle in a particle-in-a-box system with particles of the same size is the inverse of its 
collisional probability, i.e.\
\begin{equation}
t_{\rm coll}(s) = P_{\rm coll}(s)^{-1} = \left[n(s) {v}(s) \pi s^2\right]^{-1}\;,
\end{equation}
where $n(s)$ is the number density of the particles. Assuming a single particle size, we analytically determine an estimate 
for the collisional timescale of a particle as a function of its size. The flux emitted by a single nanograin will be
\begin{equation}
f(\nu;s)[{\rm Jy}] = 10^{26}\frac{\pi\lambda^2}{D^2 c}s^2 Q_{\rm abs}(s) B_{\rm RJ}\;.
\end{equation}

\noindent
where $Q_{\rm abs}$ is the absorption (emission) efficiency of the grain, $B_{\rm RJ}$ is 
the Rayleigh-Jeans approximation to a blackbody, and $D$ is the distance from Earth.
As an example, a fit to the $Q_{\rm abs}$ curves of carbon nanograins in size up to $\sim 100$ nm and for 
wavelengths of 1 - 10 $\mu$m yields 
\begin{equation}
Q_{\rm abs}(\lambda = 1-10~\micron) \approx 3\left(\frac{s}{\micron}\right)\left(\frac{\lambda}{\micron}\right)^{-1.5}\;.
\end{equation}
Therefore, the near-IR emission from a single nanograin can be written as
\begin{equation}
f(\nu;s)[{\rm Jy}] = 6\times10^{23}\frac{\pi k_{\rm b} T}{D^2}\frac{s^3}{\lambda^{3.5}}\;.
\label{eq:flux}
\end{equation}
We assume the nanograins to be super-heated to 2000 K once they are created. The number 
density can be expressed as
\begin{equation}
n(s) = \frac{F(\nu)}{f(\nu;s)V(s)}\;,
\end{equation}
where $F(\nu)$ is the total amount of excess emission. The collisional volume is calculated as
\begin{equation}
V(s) = \frac{\pi h}{2}\left[r_{\rm sub}^2-\left(r_{\rm sub}-r_{\rm gyro}\right)^2\right]\left[r_{\rm sub}+\left(r_{\rm sub}-r_{\rm gyro}\right)\right]\;,
\end{equation}
where $h$ is the disk aspect ratio, which we take to be 0.2. In the left panel of Figure \ref{fig:tcoll}, we plot the 
collisional timescales of single particles calculated using this simple analytic assumption as a function of assumed 
single grain sizes for some of the well known systems with hot excess. We plot the assumed system parameters in 
Table \ref{tab:obs}. The gyroradii of the particles and their collisional velocities (orbital velocity reduced) were 
calculated with our numerical model. The largest nanograins have the longest survival timescales, of order a few months 
to even a few years in each system.

\subsection{Required mass transport rate}

Here, we investigate the total mass transport necessary to yield detectable levels of 2 micron excess
emission. For this, we need to calculate the equilibrium particle density resulting from a continuous 
mass transport and collisional destruction. Since the magnetic field trapping timescale (i.e., in the 
absence of collisions) is many orders of magnitude longer than the collisional destruction timescale, 
we do not include it in the following calculation.

The rate at which characteristic sized particles are transported into the inner regions will be
\begin{equation}
{\tau}_{\rm prd} = \frac{\dot{M}}{m(s)}\;,
\end{equation}
while the collisional rate is expressed by
\begin{equation}
\tau_{\rm coll} = \frac{N(s)^2}{V(s)^2}{v}(s)\pi s^2 V(s)\;.
\end{equation}

\noindent
where $N(s)$ is the total number of particles of radius $s$. 
Using the flux estimates of single size particles in Equation (\ref{eq:flux}) and the collisional volume
and velocity determined by the numerical models, the necessary mass transport rate to yield a detectable 
level of steady-state particles is
\begin{equation}
\dot{M} = \left[\frac{F(\nu,s)}{f(\nu,s)}\right]^2 \frac{4 v(s) \pi^2 s^5 \rho}{3V(s)}\;.
\end{equation}
We plot the mass transfer estimates using single particle sizes in 
Figure \ref{fig:tcoll} for the systems listed in Table \ref{tab:obs}. The required mass transfer for 
the systems is in agreement with the analytic estimates we calculated in the previous section, i.e., 
under our assumptions the expected mass transport is adequate to account for required amount of hot dust.

\begin{deluxetable}{lrrrrrrrr}
\tablewidth{250pt}
\tablecolumns{9}
\tabletypesize{\tiny}
\tablecaption{Adopted stellar properties for mass transport calculations
\label{tab:obs}}
\tablehead{
\colhead{Star} & \colhead{$M_{\ast}$}    & \colhead{$R_{\ast}$}    & \colhead{$L_{\ast}$}    & \colhead{$r_{\rm sub}$} & \colhead{$T_{\rm rot}$} & \colhead{$D$}  & \colhead{$F$} & \colhead{$S_{\rm{max}}$} \\
\colhead{    } & \colhead{($M_{\odot}$)} & \colhead{($R_{\odot}$)} & \colhead{($L_{\odot}$)} & \colhead{(AU)}          & \colhead{(h)}           & \colhead{(pc)} & \colhead{(Jy)} & \colhead{(nm)}
}
\startdata
Vega           & 2.13               & 2.36             &  40       & 0.30     & 12.5          & 7.67            & 8.17  &   60  \\
$\zeta$ Aql    & 2.37               & 2.27             &  40       & 0.30     & 8.7           & 25.45           & 0.79  &  64\\
$\beta$ Leo    & 1.78               & 1.73             &  15       & 0.18     & 16.3          & 11.00           & 1.09  &  38  \\
Altair         & 1.79               & 1.80             &  11       & 0.16     & 8.9           & 5.12            & 16.48  &  53  \\
$\beta$ Pic    & 1.75               & 1.80             &  9        & 0.15     & 16.8          & 19.44           & 0.21  &  40  \\
\enddata
\end{deluxetable}

\section{Asteroids and Comets}
\label{sec:comets}

As is often the case for detectable debris disk components \citep{wyatt2005}, PRD may work too slowly to 
bring adequate material inward to account for the brightest cases of hot dust emission. A more rapid, although 
less well constrained, transport process is through asteroids and comets that are deflected into orbits that bring 
them close to the star - for brevity, we will refer to both as ``comets.'' The above discussion illustrates the 
physics involved in grain creation, trapping, and survival for this case also. As such objects disintegrate, they 
leave trails of grains and larger fragments along their Keplerian orbits. There can be a significant population of 
nanograins in comet ejecta possibly originally from the interstellar medium \citep[e.g.,][]{mcdonnell1987, greenberg1990}, 
and more nanograins will be produced as the larger particles are eroded by sublimation. As soon as the nanograins are 
exposed to the radiation of our reference A-star, they will become strongly charged electrically and will be deflected 
from the Keplerian orbits of the comet and will orbit with the magnetic field. Therefore, they will be trapped in orbits 
similar to those discussed in the previous section. This process will be expedited because many of the ``larger'' grains 
are likely to be very poorly consolidated clusters of nanograins \citep[e.g.,][]{greenberg1990,dominik2007,sanders2012} 
that will fall apart in the harsh environment close to the star. However, an important distinction compared with the 
discussion in the previous section is that their orbits can originate from a much broader range of distances from the 
star, i.e., from any distance to which a larger body has survived. Therefore, there will be a lower space density for 
a given mass of nanograins, and the destruction rate by collisions will be lower.

To evaluate the requirements for this transport mechanism, we assume that the nanograins have a typical lifetime of 
four months, that the comet producing them has a density of 1 g cm$^{-3}$, that 7\% of the comet mass is already 
\citep{mcdonnell1987,sanders2012} or is converted to nanograins, and that an equilibrium mass of $10^{15}$ kg of 
nanograins is required to produce the hot excess (Section \ref{sec:exploration}). The destruction of three comets a 
year, each of 15 km radius, is then sufficient to produce a hot excess of $\sim 1\%$ at 2 $\mu$m around our 
reference A-star. Thus, plausible additional mass transport in this manner can augment that from PRD and make the 
models for the hot dust more robust.

\section{Testable predictions of the model}
\label{sec:tests}

There are three results of our modeling that can be tested: 1.) in Section \ref{sec:exploration} 
we derived some general requirements imposed by the observed roughly Rayleigh-Jeans spectrum of the hot 
excesses; 2.) in Section \ref{sec:rotation}, we found that rapidly rotating stars should be able to retain larger 
equilibrium populations of nanograins than slow rotators; and 3.) in Section \ref{sec:fieldstrength} we pointed out 
that very large magnetic fields would result in rapid nanograin destruction and 
that one should therefore not necessarily expect 
stars with such fields to have large excesses. In this third case, since the type of A-star with large magnetic fields also 
rotates slowly, we can only test the effect of the two variables together. We examine each of these predictions in turn.

\subsection{Spectrum and size of emitting regions}

We deduced in a general way in Section \ref{sec:exploration} that the emitting grains could not be larger than about 
200 nm, to avoid producing too much radiation at 10 $\mu$m. In Section 4.4 we derived an upper limit of 
121 nm for the size of grains that could be trapped magnetically around our reference A-star. It is inherent in the 
general behavior of the hot excesses that the grains must be close to the star, and this has been confirmed directly for 
Vega \citep{mennesson2011}. By trapping nanograins as they are released near 0.2 AU due to sublimation or breakdown of 
comets, this requirement is automatically satisfied by our hypothesis for the hot excesses. 

\subsection{Rotation rates and hot excesses}
\label{sec:rothot}

We now consider the prediction of an association of hot excesses with rapidly rotating stars. To compare rotation rates, 
we used vsini\footnote{vsini's for HIP 21589, 27288, 27321, 41307, 53910, 54872, 57632, 59774, 60965, 61622,  87108, 88771, 
93747, 94114,  97649, 98495, \& 105199 are from \citet{zorec2012}, for HIP 50191 \& 71908 from \citet{glebocki2005}, for 
HIP 2072, 11001, 102333, \& 109285 from \citet{vanbelle2012}, for HIP 71284 from \citet{royer2002}, and for HIP 74824 
from \citet{diaz2011}.}. Although rotation determination from this parameter is uncertain for any individual star due 
to unknown inclination effects, a large value is an unequivocal indication of fast rotation, allowing us to compare stars 
known to rotate rapidly with a mixed control sample. We confined the sample to 
spectral types of F2V and earlier (and excluded giants), since spin down is a dominant issue otherwise. We found that the 
average vsini for stars in \citet{absil2013} and \citet{ertel2014} without detected hot excesses is 
$141 \pm 20$ km s$^{-1}$, where we quote the error of the mean based on the rms scatter of the measured values for stars 
without hot excesses. Similarly, vsini for the stars with detected excesses is $210 \pm 27$ km s$^{-1}$, with the error 
quoted for the mean and based on the rms scatter of the measured values for the stars with hot excesses\footnote{vsini $\sim$ 200 km s$^{-1}$ corresponds to a rotational period of $\sim$ 10 hours for a star viewed from the equatorial direction.}. The difference is 
$69 \pm 34$ km s$^{-1}$, that is at 2$\sigma$ significance, or with a 2.3\% probability of being due to chance (given that 
the difference is in the predicted direction). We did not include Vega in any of these averages because it has a small vsini 
but is known to be a very rapid rotator. However, it adds to the evidence that hot excesses are associated with rapid 
rotation. Therefore, the prediction that hot excesses would tend to be found in rapidly rotating stars is supported.

\subsection{Excesses in Ap/Bp stars}  

Naively, if a little magnetic field is good for hot excesses, one might expect that a lot of magnetic field would be better. 
However, our modeling indicates that this expectation may not be fulfilled: stars with strong magnetic fields will confine 
nanograins more tightly, resulting in a higher rate of collisions and more rapid nanograin destruction.  
We used a sample of Ap stars to test this possibility. Many of the chemically peculiar Ap stars 
have measured fields of tens to hundreds of Gauss, and it is believed that such fields are a 
general characteristic of the class \citep[e.g.,][]{auriere2007, mathys2009, braithwaite2014, kochukhov2014}. 
However, these stars are also slow rotators \citep{tassoul2004}, so they provide a mixed test for the effects of 
magnetic fields and of rotation. 

To look for excesses around Ap stars, we started with the \citet{mcdonald2012} 
list of stars. We accepted only those members with modeled temperatures between 6500 K and 13,000 K. We drew an initial 
sample of Ap and Bp star candidates on this list from \citet{renson2009}. We flagged all likely Ap, Am, or HgMn stars and 
then used the information in \citet{skiff2014} to eliminate all but the Ap and Bp ones. 
There are $\sim$ 460 Ap and Bp stars in the \citet{mcdonald2012} study with usable 2MASS photometry. For a 
control sample, we accepted all 19473 non-Ap stars in \citet{mcdonald2012} meeting our input citeria. 
We found no evidence for hot excesses in the $J - K_S$ colors between 
the Ap stars and the control sample.
The large magnetic fields in Ap/Bp stars do not appear to result in a high incidence of hot excesses. 
This result may arise because the 
effect of slow rotation in these stars overcomes any trend toward magnetic nanograin trapping.

In fact, the average $J - K_S$ color of the Ap stars relative to the trend for normal A stars is slightly (-0.008 mag) negative,
 which may be a result of subtle 
spectral differences due to their anomalous abundances. 
Another possibility is their slow 
rotation \citep{tassoul2004} and the resulting relatively uniform surface temperatures. That is, rapid rotation 
will establish a temperature gradient over the surface of the star, with the equatorial region cooler than the 
polar ones because of its larger radius. Spectral typing is done in the optical, where the hot polar regions will 
be relatively prominent. For A-stars, the $J-K_S$ color is largely in the Rayleigh-Jeans regime, where the contribution 
of the cooler equatorial regions will be relatively greater than in the optical. Thus, a sample 
of stars that includes many rapidly rotating ones might have a $J-K_S$ colors redder than implied by the spectral 
types, whereas a sample of slowly rotating stars should have $J-K_S$ matching the spectral types more closely.

\subsection{Rotation and infrared colors}

The possibility that rapid rotating stars might show anomalously red infrared colors for their spectral types would
undermine the argument of slightly red $J-K$ for hot excess stars that we made in Section \ref{sec:hottest} and for 
slow rotation contributing to the lack of such an excess for ApBp stars just above. Vega can 
be taken as an extreme example, since it is rotating nearly at the breakup rate and is virtually perfectly pole-on, 
maximizing the exposure to the hot poles and resulting in the assignment of a relatively early spectral type 
\citep{aufdenberg2006, castaneda2014}. \citet{rieke2008} estimate that as a result $J-K_S$ is 0.01 mag redder than 
expected for the assigned spectral type; given the extreme nature of Vega, an effect of $1 - 2\%$ is an upper limit 
to what might occur with other stars. 

We used the \citet{mcdonald2012} sample to test for this effect. We took all members of the sample with vsini 
$\ge$ 200 km s$^{-1}$ and compared their $J - K_S$ with the sample members with smaller vsini, taking rotation measures 
from \citet{glebocki2005, zorec2012}. The comparison is made difficult by the relatively large number of 1.) rapidly 
rotating emission-line stars, which need to
be excluded because of the potential of K-band excesses due to free-free emission, 2.) stars with faint red companions, 
which were reduced by eliminating those detected in the X-ray \citep{derosa2011}, and 3.) binaries in general. The latter 
category is troublesome because tidal locking in close binaries can sustain fast rotation and hence bias the sample due
to undetected close companions. To eliminate such cases, we rejected all examples with $J-K_S \ge 0.08$, on the basis that 
this value is substantially larger than the $1 - 2\%$ possible solely from fast rotation and indicates an excess due to
another cause. We also rejected the few cases with $J-K_S \le -0.08$. 

The average $J-K_S$ for the stars with the higher vsini was larger by $0.003 \pm 0.002$ mag. In Section \ref{sec:rothot}
we found indications that hot excesses tend to be associated with large vsini. According to \citet{absil2013}, about 50\% of 
A-stars have hot excesses at typical levels of $\sim 1\%$. Therefore, the apparent tiny excess at 
$K_S$ could just arise from the expected number and extent of hot dust excesses. There is no convincing effect of
rapid rotation on the intrinsic stellar $J-K_S$ color. We therefore believe that the result in Section \ref{sec:hottest} 
is valid and that the small color difference for the Ap stars is primarily due to a spectral effect arising from their 
anomalous abundances.

\section{Conclusions}
\label{sec:end}

We have studied the hot near-infrared excesses detected by interferometric measurements, particularly of A-stars. 
Our main conclusions are:

\begin{enumerate}

\item Infrared photometry shows a slightly (1.1\%) redder \mbox{$J-K_S$} color for stars reported to have these 
excesses than for those reported not to have excesses. Although only at $\sim$ 1.5 $\sigma$ significance, 
this result supports the reality of the detections. 

\item Given the roughly Rayleigh-Jeans spectral energy distributions of the excesses, we conclude in agreement 
with previous studies that they must be generated by hot dust grains less than 200 nm in radius. Under typical 
assumptions for circumstellar dusty debris, such grains would be blown away from A-stars by radiation pressure 
force. 

\item However, the grains will rapidly acquire significant electrical charge. We demonstrate that these charged 
grains can have significant trapping times in the stellar magnetic field via Lorentz forces that place the 
particles on epicyclic orbits. Trapping occurs even for fields as small as 0.1 G, less than the field strengths 
measured for Sirius and Vega.

\item Magnetic trapping also naturally accounts for other aspects of the hot grain population; it imposes an upper 
limit on the grain sizes of just over $\sim$ 100 nm, it keeps the grains at 0.1 - 0.2 AU from the star, and it can 
maintain a sufficient steady-state population of grains to account for the hot emisison.

\item Additional predictions from the trapping model, that the hot excesses are likely to be associated with rapid 
stellar rotation and that such excesses would not be preferentially associated with very large magnetic fields, are 
also supported by observation. 

\end{enumerate}

\section{Acknowledgements}

We thank Kaitlin Kratter, Mike Sitko, Rik van Lieshout, Mark Wyatt and the anonymous referee for helpful comments. 
This work was supported 
by NASA grants NNX13AD82G and 1255094. This publication makes use of data products from the Two Micron All Sky Survey, 
which is a joint project of the University of Massachusetts and the Infrared Processing and Analysis Center/California 
Institute of Technology, funded by the National Aeronautics and Space Administration and the National Science Foundation. 
This research also has made use of the VizieR catalogue access tool, CDS, Strasbourg, France. The original description 
of the VizieR service was published in A\&AS 143, 23.

\appendix
\center
{\bf Procedures to transform heritage photometry to 2MASS}

\flushleft

We have assembled as much high-quality $J,H,K_S$ photometry as possible for bright stars. We evaluated various sources 
of heritage near-infrared photometry and accepted those that appeared to have good internal accuracy and that could be 
transformed accurately onto the 2MASS system. For this paper, we took photometry from:

\begin{enumerate}

\item
2MASS for HD 15008, 134083 \& 210049; 

\item
\citet{allen1983} for HD 15798 \& 71155; 

\item
\citet{aumann1991} (which we found to be consistent with the CIT system) for HD 2262*, 15008, 30652, 38678, 39060*, 102647*, 104731*, 106591, 128167, 134083, 135379, 139664, 142860, 164259, \& 210302*; 

\item
\citet{carter1990} for HD 2262*, 15798, 39060*, 71155, 88955, 97603, 108767*, 109787, 129502, \& 210049; 

\item
the ESO standards list \citep{bouchet1991} for HD 2262*, 15008, 39060*, 71155, 88955, 109787, 135379, 178253, 197692, \& 210049;

\item
\citet{glass1974} for HD 15798, 39060*, 71155, 97603, 108767*, 109787, 129502;

\item
\citet{johnson1966} (where we accepted only stars measured at least three times under good conditions) for HD 22484, 30652, 48737, 88955, 95418, 97603, 102647*, 104731*, 106591, 108767*, 128167, 129502, 134083, 142860, 161868, 177724*, 187642*, \& 203280*;  

\item
\citet{kidger2003} for HD 29388 \& 187642*;

\item
\citet{kimeswenger2004} for HD 102647*;

\item
\citet{mcgregor1994} for HD 15798, 71155; 

\item
Su, K. Y. L. \& Rieke, G. H. (unpublished) for HD 3302, 22001, 135379, 160032, 188228, 197157, 197692, 210302*, \& 213845 (SAAO system);  

\item
and the UKIRT standards list for HD 102647* \&  128167.

\end{enumerate}

The stars with hot excesses are indicated by * and can be seen to be distributed similarly to the other stars, e.g., their 
photometry should be homogeneous with that for stars without hot excesses. 

\bigskip

We used transformations from \citet{carpenter2003} to convert photometry from the Anglo-Australian Observatory, 
CIT, ESO, and UKIRT systems to the 2MASS one. To convert SAAO to 2MASS we used \citet{koen2007}.
We also derived transformations as follows:

\begin{itemize}

\item
K(2MASS) = K(Johnson)-0.0567+0.056(J-K)

\item
J(2MASS)= J(Johnson) - 0.69 + 0.0285(J-K)

\item
K(2MASS)= K(Kidger)+0.008+0.0257(J-K)

\item
J(2MASS)= J(Kidger)-0.0199+0.0483(J-K)

\item
K(2MASS) = K(Kimeswenger) -.006

\item
J(2MASS) = J(Kimeswenger) 

\end{itemize}

It is difficult to estimate errors for transformed photometry. A number of tests indicated that the products have similar errors as the 
2MASS photometry and track the 2MASS colors well. Therefore, where there were multiple measurements of a star, we averaged them 
with equal weights after transforming them onto the 2MASS system.

%


\begin{thebibliography}{}

\bibitem[Absil et al. (2006)]{absil2006}
Absil, O., di Folco, E., M\'erand, A. et al. 2006, A\&A, 452, 237

\bibitem[Absil et al. (2008)]{absil2008}
Absil, O., di Folco, E., M\'erand, A. et al. 2008, A\&A, 487, 1041

\bibitem[Absil et al. (2013)]{absil2013}
Absil, O., Defr\`ere, D, Coud\'e du Foresto, V., Di Folco, E. et al. 2013, A\&A, 555, 104

\bibitem[Akeson et al. (2009)]{akeson2009}
Akeson, R. L., Ciardi, D. R., Millan-Gabet, R. et al. 2009, ApJ, 691, 1896

\bibitem[Alina et al. (2012)]{alina2012}
Alina, D., Petit, P., Ligni\`eres, F., et al. 2012, AIPC, 1429, 82

\bibitem[Allen \& Cragg (1983)]{allen1983}
Allen, D. A. \& Cragg, T. A. 1983, MNRAS, 203, 777


\bibitem[Aufdenberg et al. (2006)]{aufdenberg2006}
Aufdenberg, J. P., M\'erand, A., Coud\'e du Foresto, V. et al. 2006, ApJ, 645, 664


\bibitem[Aumann \& Probst (1991)]{aumann1991}
Aumann, H. H., \& Probst, R. G. 1991, ApJ, 368, 264

\bibitem[Auri\`ere et al. (2007)]{auriere2007}
Auri\`ere, M., Wade, G. A., Silvester, J., Ligni\`enes, F. et al. 2007, A\&A, 475, 1053 

\bibitem[Babel(1995)]{babel1995}
Babel, J. 1995, A\&A, 301, 823

\bibitem[Bonsor et al. (2013)]{bonsor2013}
Bonsor, A., Raymond, S. N., \& Augereau, J.-C. 2013, MNRAS, 433, 2938

\bibitem[Bonsor et al. (2014)]{bonsor2014}
Bonsor, A., Raymond, S. N., Augereau, J.-C., \& Ormel, C. W. 2014, MNRAS, 441, 2380

\bibitem[Bouchet et al. (1991)]{bouchet1991}
Bouchet, P., Schmider, F. X., \& Manfroid, J. 1991, A\&AS, 91, 409

\bibitem[Braithwaite \& Cantiello (2013)]{braithwaite2013}
Braithwaite, J., \& Cantiello, M. 2013, MNRAS, 428, 2789

\bibitem[Braithwaite (2014)]{braithwaite2014}
Braithwaite, J. 2014, Proc. IAU Symp., 302, 255

\bibitem[Burns et al. (1979)]{burns1979}
Burns, J. A., Lamy, P. L., \& Soter, S. 1979, Icarus, 40, 1

\bibitem[Carpenter (2003)]{carpenter2003}
Carpenter, J. M. 2003, http://www.astro.caltech.edu/~jmc/2mass/v3/transformations/

\bibitem[Carter (1990)]{carter1990}
Carter, B. S. 1990, MNRAS, 242, 1

\bibitem[Casagrande et al. (2010)]{casagrande2010}
Casagrande, L., Ramirez, I., Mel\'endez, J., Bessell, M., \& Asplund, M. 2010, A\&A,512, 54

\bibitem[Casta\~neda et al. (2014)]{castaneda2014}
Casta\~neda, D., Deupree, R. G., \& Aufdenberg, J. P. 2014, astroph 1411.1673

\bibitem[De Rosa et al. (2011)]{derosa2011}
De Rosa, R. J. Bulger, J., Patience, J. et al. 2011 MNRAS, 415, 854

\bibitem[Diaz et al. (2011)]{diaz2011}
Diaz, C. G., Gonzalez, J. F., Levato, H., \& Grosso, M. 2011, A\&A, 531, 143

\bibitem[Defr\`ere et al. (2011)]{defrere2011}
Defr\`ere, D., Absil, O., Augereau, J.-C. et al. 2011, A\&A, 534, 5

\bibitem[De Rosa et al. (2011)]{derosa2011}
De Rosa, R. J., Bulger, J., Patience, J. et al. 2011, MNRAS, 415, 845

\bibitem[Dominik et al. (2007)]{dominik2007}
Dominik, C., Blum, J., Cuzzi, J. N., \& Wurm, G. 2007, Protostars \& Planets, page 783. 

\bibitem[Ertel et al. (2014)]{ertel2014}
Ertel, S., Absil, O., Defr\`ere, D. et al. 2014, A\&A, 570, 128

\bibitem[Fedkin et al. (2006)]{fedkin2006}
Fedkin, A. V., Grossman L., \& Ghiorso, M. S. 2007, Geochimica et Cosmochimica Acta, 70, 206


\bibitem[Glass (1974)]{glass1974}
Glass, I. S. 1974, MNSSA, 33, 53

\bibitem[Glebocki \& Gnacinski (2005)]{glebocki2005}
Glebocki, R. \& Gnacinski, P. 2005, ESA SP-5600, 571

\bibitem[Greenberg \& Hage (1990)]{greenberg1990}
Greenberg, J. M., \& Hage, J. I. 1990, ApJ, 361, 260




\bibitem[Ignatov (2009)]{ignatov2009}
Ignatov, A. M. 2009, Plasma Physics Reports, 35, 647

\bibitem[Johnson et al. (1966)]{johnson1966}
Johnson, H. L., Mitchell, R. I., Iriarte, B., \& Wisniewski, W. Z. 1966, Comm. Lun. Plan. Lab., 4, 99

\bibitem[Kazenas \& Tsevtkov (2008)]{kazenas2008}
Kazenas, E. K., \& Tsvetkov, Yu. V. 2008, "Thermodynamics of the Evaporation of Oxides"

\bibitem[Kidger \& Mart\'in-Luis (2003)]{kidger2003}
Kidger, M. R. \& Mart\'in-Luis, F. 2003, AJ, 125, 3311

\bibitem[Kimeswenger et al. (2004)]{kimeswenger2004}
Kimeswenger, S., Lederle, C., Richichi, A. et al. 2004, A\&A, 413, 1037

\bibitem[Kobayashi et al. (2009)]{kobayashi2009}
Kobayashi, H., Watanabe, S.-I., Kimura, H., \& Yamamoto, T. 2009, Icarus, 201, 395

\bibitem[Kobayashi et al. (2011)]{kobayashi2011}
Kobayashi, H., Kimura, H., Watanabe, S.-I.,  Yamamoto, T., \& M\"uller, S. 2011, Earth, Plan., \& Spc, 63, 1067

\bibitem[Kochukhov (2014)]{kochukhov2014}
Kochukhov, O. 2014, "Putting A Stars into Context: Evolution, Environment, and Related Stars," Publishing House "Pero", pp. 389-397

\bibitem[Koen et al. (2007)]{koen2007}
Koen, C., Marang, F., Kilkenny, D., \& Jacobs, C. 2007, MNRAS, 380, 1433

\bibitem[Lamoreaux et al. (1987)]{lamoreaux1987}
Lamoreaux, R. H., Hildenbrand, D. L., \& Brewer, L. 1987, J. Phys. Chem. Ref. Data, 16, No. 3

\bibitem[Lebreton et al. (2013)]{lebreton2013}
Lebreton, J., van Lieshout, R., Augereau, J.-C. et al. 2013, A\&A, 555, 146

\bibitem[Lefevre (1975)]{lefevre1975}
Lefevre, J. 1975, A\&A, 41, 437

\bibitem[Ligni\`eres et al. (2009)]{lignieres2009}
Ligni\`eres, F., Petit, P., B\"ohm, T., \& Auri\`ere, M. 2009, A\&A, 500, 41

\bibitem[Liou \& Zook (1999)]{liou1999}
Liou, J.-C., \& Zook, H A. 1999, AJ, 118, 580

\bibitem[Ma et al. (2013)]{ma2013}
Ma, Q., Matthews, L. S., Land, V., \& Hyde, T. W. 2013, ApJ, 763, 77

\bibitem[Mann et al. (2006)]{mann2006}
Mann, I., K\"ohler, M., Kimura, H., Czechowski, A., \& Minato, T. 2006, Astron. Astrophys. Rev., 13, 159

\bibitem[Mann et al. (2007)]{mann2007}
Mann, I., Murad, E., \& Czechowski, A. 2007, Plan. \& Space Sci., 55, 1000

\bibitem[Mann et al. (2014)]{mann2014}
Mann, I., Meyer-Vernet, N., \& Czechowski, A. 2014, Physics Reports, 536, 1

\bibitem[Mathys (2009)]{mathys2009}
Mathys, G. 2009, ASP Conf. Ser. 405, 473

\bibitem[Mawet et al. (2011)]{mawet2011}
Mawet, D., Mennesson, B., Serabyn, E., Stapelfeldt, K., \& Absil, O. 2011, ApJL, 738, 12

\bibitem[McDonald et al. (2012)]{mcdonald2012}
McDonald, I., Zijlstra, A. A., \& Boyer, M. L. 2012, MNRAS, 427, 343

\bibitem[McDonnell et al. (1987)]{mcdonnell1987}
McDonnell, J. A M., Alexander, W. M., Burton, W. M. et al. 1987, A\&A, 187, 719

\bibitem[McGregor (1994)]{mcgregor1994}
McGregor, P. J. 1994, PASP, 106, 508

\bibitem[Mennesson et al. (2011)]{mennesson2011}
Mennesson, B., Serabyn, E., Hanot, C, Martin, S. R., Liewer, K., \& Mawet, D. 2011, ApJ, 736, 14

\bibitem[Mennesson et al. (2014)]{mennesson2014}
Mennesson, B., Milan-Gabet, R., Serabyn, E., Colavita, M. M. et al. 2014, ApJ, 797, 119



\bibitem[Ohnishi et al. (2008)]{ohnishi2008}
Ohnishi, N., Bringa, E. M., Remington, B. A., et al. 2008,J. Phys. Conf. Series, 112, 042017

\bibitem[Palik (1997)]{palik1997}
Palik, E. D. 1997, "Handbook of Optical Constants of Solids," Academic Press: Waltham, MA, USA

\bibitem[Papoular (2004)]{papoular2004}
Papoular, R. 2004, A\&A, 414, 573

\bibitem[Parker (1958)]{parker1958}
Parker, E. N. 1958, ApJ, 128, 664

\bibitem[Parker (1964)]{parker1964}
Parker, E. N. 1964, ApJ, 139, 951


\bibitem[Pedersen \& G\'omez de Castro (2011)]{pedersen2011}
Pedersen, A., \& G\'omez de Castro, A. I. 2011, ApJ, 740, 77

\bibitem[Petit et al. (2011)]{petit2011}
Petit, P., Ligni\`eres, F., Auri\`ere, M., Wade, G. A. et al. 2011, A\&A, 532L, 13

\bibitem[Pickles \& Depagne (2010)]{pickles2010}
Pickles, A., \& Depagne, \'E. 2010, PASP, 122, 1437

\bibitem[Raymond \& Bonsor (2014)]{raymond2014}
Raymond, S. N. \& Bonsor, A. 2014, MNRAS, 442L, 18

\bibitem[Renson \& Manfroid (2009)]{renson2009}
Renson, P., \& Manfroid, J. 2009, A\&A, 498, 961

\bibitem[Rieke et al. (2008)]{rieke2008}
Rieke, G. H., Blaylock, M., Decin, L. et al. 2008, AJ, 135, 2245

\bibitem[Royer et al. (2002)]{royer2002}
Royer, F., Grenier, S., Baylac, M.-O., G\'omez, A. E., \& Zorec, J. 2002, A\&A, 393, 897

\bibitem[Sanders \& Velbel (2012)]{sanders2012}
Sanders, N. E., \& Velbel, M. A. 2012, Meteroitics \& Plan. Sci., 47, 594

\bibitem[Schatten et al. (1969)]{schatten1969}
Schatten, K. H., Wilcox, J. M., \& Ness, N. F. 1969, Solar Phys., 6, 442

\bibitem[Skiff (2014)]{skiff2014}
Skiff, B A. 2014, VizieR Online Data Catalog of Stellar Spectral Types

\bibitem[Su \& Rieke (2014)]{su2014}
Su, K. Y. L., \& Rieke, G. H. 2014, IAU Symp., 299, 318 

\bibitem[Tassoul (2004)]{tassoul2004}
Tassoul, J.-L. 2004, "Stellar Rotation," Cambridge University Press: Cambridge, UK, page 192

\bibitem[UKIRT (2000)]{ukirt2000}
UKIRT bright standards available at http://www.gemini.edu/sciops/instruments/near-ir-resources/nir-photometric-standard-stars/ukirt-bright-standards

\bibitem[van Belle (2012)]{vanbelle2012}
van Belle, G. T. 2012, Ast. \& Astrophys. Rev., 20, 51

\bibitem[van Lieshout et al. (2014)]{vanlieshout2014}
van Lieshout, R., Dominik, C., Kama, M., \& Min, M. 2014, A\&A, 571, 51

\bibitem[Voges et al. (1999)]{voges1999}
Voges, W., Aschenbach, B., Boller, Th. et al. 1999, A\&A, 349, 389

\bibitem[Weingartner \& Draine (2001)]{weingartner2001}
Weingartner, J., \& Draine, B. 2001, ApJS, 134, 263

\bibitem[Wolf (2015)]{wolf2015}
Wolf, S. 2015, http://www1.astrophysik.uni-kiel.de/dds/

\bibitem[Wurz (2012)]{wurz2012}
Wurz, P. 2012, in "Nanodust in the Solar System: Discoveries and Interpretations," ed. I Mann, 
N. Meyer-Vernet, A. Czechowski, Springer: Heidelberg, pp 161-178

\bibitem[Wyatt (2005)]{wyatt2005}
Wyatt, M. C., 2005, A\&A, 433, 1007

\bibitem[Wyatt (2008)]{wyatt2008}
Wyatt, M. C. 2008, \araa, 46, 339

\bibitem[Zorec \& Royer (2012)]{zorec2012}
Zorec, J., \& Royer, F. 2012, A\&A, 537, 120

\end{thebibliography}
\end{document}